\documentclass[%
reprint,
superscriptaddress,
nobibnotes,
amsmath,amssymb,
aps,
pra,
]{revtex4-1}

\usepackage{graphicx}
\usepackage{dcolumn}
\usepackage{bm}
\usepackage{hyperref}
\hypersetup{
    breaklinks=true,
    bookmarks=true,
    pdfauthor={},
    pdftitle={},
    colorlinks=true,
    citecolor=blue,
    urlcolor=blue,
    linkcolor=magenta,
    pdfborder={0 0 0}
}

\usepackage{mathtools} 
\usepackage{upgreek} 

\usepackage{color,soul} 
\usepackage[normalem]{ulem} 
\usepackage{xfrac} 
\usepackage{braket} 
\usepackage{cases} 
\usepackage{placeins} 
\usepackage{booktabs} 

\usepackage{floatrow} 
\usepackage[caption=false]{subfig}  

\usepackage{xr} 
\begin{document}

\floatsetup[figure]{style=plain,subcapbesideposition=top} 
\floatsetup[table]{style=plain, capposition=top} 

\preprint{APS/123-QED}

\title{First-principles studies of Schottky barriers and tunneling properties at Al(111)/Si(111) and CoSi$_2$(111)/Si(111) interfaces}


\author{J. K. Nangoi}
\email[Corresponding author: nangoi@ucsb.edu]{}
\affiliation{
    Materials Department, University of California, Santa Barbara, California 93106-5050, USA
}

\author{C. J. Palmstr{\o}m}
\affiliation{
    Materials Department, University of California, Santa Barbara, California 93106-5050, USA
}

\author{C. G. Van de Walle}
\affiliation{
    Materials Department, University of California, Santa Barbara, California 93106-5050, USA
}


%


\begin{abstract}

    We present first-principles calculations of Schottky barrier heights (SBHs) at
    interfaces relevant for silicon-based merged-element transmon qubit devices. 
    Focusing on Al(111)/Si(111) and CoSi$_2$(111)/Si(111), we consider various possible interfacial structures, for which we study the relaxations of the atoms near the interface, calculate the formation energies and Schottky barrier heights, and provide estimates of the Josephson critical currents based on the WKB tunneling formalism as implemented in the Simmons/Tsu-Esaki model. 
    We find that the formation energies and SBHs are very similar for all Al(111)/Si(111) structures, yet vary significantly for the CoSi$_2$(111)/Si(111) structures.  We attribute this to the more covalent character of bonding at CoSi$_2$/Si, which leads to configurations with distinct atomic and electronic structure.
    Our estimated Josephson critical currents, which govern the behavior of merged-element transmons, provide insight into the trends as a function of Schottky-barrier height.
    We show that desirable qubit frequencies of 4--5~GHz can be obtained with a Si barrier thickness of about 5--10~nm, and demonstrate that the critical current density as a function of Schottky barrier height can be modeled based on the tunneling probability for a rectangular barrier.
    
\end{abstract}


\maketitle



\section{Introduction} \label{sec:intro}

Al/Si interfaces have been studied since the 1970s for applications in electronic devices~\cite{ref:louie, ref:tejedor, ref:legoues, ref:yapsir, ref:miura_typeB, ref:miura_typeA, ref:fortuin, ref:zavodinsky, ref:skachkov} 
and to understand and control structural properties of Al/Si cast alloys for use in automotive and aerospace~\cite{ref:liu, ref:wang, ref:wu}. 
Recently, Al/Si interfaces have been used in 
novel qubit devices called merged-element transmons (MET)~\cite{ref:finmet}. 
Transmons, the standard qubits for quantum computing in superconducting circuits, are conventionally  based on metal/oxide/metal Josephson junctions and 
paddle capacitors with a large footprint.
It has been proposed that scalability can be significantly improved by replacing the large external shunt capacitor of a traditional transmon with the intrinsic capacitance of the Josephson junction~\cite{ref:met_aSi}.

Replacing the oxide with a lower-band-gap material such as Si allows for significantly thicker tunnel barriers in the Josephson junction, leading to smaller variations in the Josephson current densities than those observed in AlO$_x$ MET devices \cite{ref:alox, ref:met_alox, ref:finmet}.
While the first MET devices were based on amorphous Si~\cite{ref:met_aSi}, use of float-zone crystalline Si should additionally reduce dielectric losses and minimize the formation of two-level system spectral features.
The significantly smaller size compared to traditional transmons should allow for scalable fabrication, especially when combined with fin fabrication and atomic-layer or digital etching in the so-called FinMETs~\cite{ref:finmet}. 

The enhanced control over the junction between the superconducting metal and the tunnel barrier, which is at the heart of the performance improvements, prompts renewed scrutiny of the properties of this interface.
In the present work we focus on interfaces between the metal and the Si(111) surfaces that constitute the sides of the fins in a FinMET~\cite{ref:finmet}.
For the metal, we consider Al, which is widely used in superconducting qubits~\cite{ref:met_alox}. 
In addition to Al, 
CoSi$_2$ could be an excellent choice as the metal in a MET because CoSi$_2$ is superconducting with a $T_c$ of 1.26 K~\cite{ref:cosi2-superconducts}, is lattice-matched to Si within $\sim$2\%~\cite{ref:si_exp-lattConst, ref:cosi2-exp-latt-const}, and has been demonstrated to grow epitaxially on Si(111)~\cite{ref:tung-1982_cosi2-epitaxial-growth, ref:sirringhaus, ref:meyer, ref:fang-2021_cosi2-epitaxial-growth}. 

In order to elucidate the atomic and electronic structure of these metal/semiconductor interfaces we perform first-principles calculations based on density-functional theory (DFT) with a hybrid functional.
A key quantity is the Schottky barrier height (SBH), which determines 
the current flowing from the metal to the semiconductor. 

While there are recent first-principles studies of Al/Si interfaces~\cite{ref:wang, ref:skachkov}, none of them considered Al(111)/Si(111), the focus of the present work. 
Previous calculations of Al(111)/Si(111) interfaces were limited by various approximations and assumptions. 
An early study by Louie {\it et al.}~\cite{ref:louie} treated the Si atomistically but approximated the Al layer with a jellium potential. 
Another early study by \citet{ref:tejedor} incorporated some atomistic corrections to the jellium potential. 
Zavodinsky and Kuyanov~\cite{ref:zavodinsky} used DFT with the local-density approximation (LDA) and the pseudopotential method to treat the Al(111)/Si(111) interface system atomistically, but they forced the in-plane periodicity of Al to match that of the Si substrate, introducing an unrealistically large in-plane strain of $\sim$35\% to the Al layers. 
Information about the in-plane periodicity is actually available from experiment~\cite{ref:legoues, ref:yapsir, ref:fortuin}, revealing that four Al repeat units match to three Si repeat units. 
Finally, \citet{ref:wu} considered this realistic interfacial structure, but used molecular dynamics with empirical interatomic potentials
and did not calculate the SBHs. 
Our present work uses realistic interfacial structures, as well as a state-of-the-art hybrid functional to produce reliable electronic structure properties.

For CoSi$_2$(111)/Si(111), six different structures have been proposed by previous theoretical and experimental studies~\cite{ref:gibson, ref:tung-1984, ref:hamann, ref:catana, ref:bulle, ref:stadler, ref:li, ref:seubert} based on three different coordination numbers of the Co atom at the interface (5, 7, or 8) and two types of the stacking of the CoSi$_2$ layers relative to the Si layers (A or B), explained in more detail in Sec.~\ref{subsec:results-cosi2_struct-en}. Following the notation used in Ref.~\cite{ref:stadler}, we call these structures A5, B5, A7, B7, A8, and B8. 
For these structures, a number of first-principles SBH calculations have been reported. 
\citet{ref:stadler} used a generalized-gradient approximation (GGA) functional~\cite{ref:gga-b,ref:gga-p} 
and considered the A7, B7, A8, and B8 structures;
they acknowledged that the thicknesses of the CoSi$_2$ layers in their calculations were too small, resulting in an uncertainty of $\sim$0.2~eV in their calculated alignments. 
\citet{ref:zhao} used the LDA functional~\cite{ref:lda-ca} 
and considered only the B8 structure.
\citet{ref:gao} also considered only the B8 structure, and used LDA for optimizing the structure and the hybrid functional of Heyd, Scuseria, and Ernzerhof (HSE)~\cite{ref:HSE03, ref:HSE06} to calculate the SBHs. 
Finally, \citet{ref:wasey} used PBE~\cite{ref:pbe} and considered the A7, B7, A8, and B8 structures.
None of these works considered the A5 and B5 structures~\cite{ref:tung-1984}.
A5 was found by \citet{ref:hamann} to have significantly higher formation energy than the structures mentioned above; however, as also discussed in~\cite{ref:stadler}, Ref.~\cite{ref:hamann} did not relax the atoms, and therefore the calculated formation energy could be overestimated. 

We also note that
Refs.~\onlinecite{ref:zhao}, \onlinecite{ref:gao}, and \onlinecite{ref:wasey} use the layer-projected density-of-states approach to calculate the SBHs.
As explained in Sec.~\ref{sec:approach}, we consider this to be less accurate than the potential-alignment method used in the present work~\cite{ref:vandewalle87, ref:stadler, ref:delaney, ref:ma, ref:weston}. 

In all our calculations we assume the silicon layer to be sufficiently thick to serve as the ``substrate'' with its lattice parameter fixed to the equilibrium bulk value.  Any strain present to accommodate the lattice mismatch is assumed to occur entirely within the metal layer.
For Al(111)/Si(111) we consider the structures suggested by experiments~\cite{ref:legoues, ref:yapsir, ref:miura_typeA, ref:miura_typeB, ref:fortuin}.
For CoSi$_2$(111)/Si(111), we consider six different structures proposed by previous theoretical and experimental studies~\cite{ref:gibson, ref:tung-1984, ref:hamann, ref:catana, ref:bulle, ref:stadler, ref:li, ref:seubert}. 
For all materials and structures considered, we report the magnitude of the relaxations of the atoms in the layers near the interface, the interplanar distances of said layers, the interface formation energies, and the SBHs. 

Using the calculated SBHs, we also report estimates on the Josephson critical current densities (which determine the transmon qubit's resonance frequency~\cite{ref:kim, ref:met_aSi}) for various silicon tunneling barrier thicknesses. 
The estimations are based on (1) the relationship between the Josephson critical current and the normal-state (nonsuperconducting) tunneling current~\cite{ref:ambegaokar, ref:tinkham}, (2) the formulation of the normal-state tunneling current through a tunnel junction by both Simmons~\cite{ref:simmons1963, ref:simmons1964} and also Tsu and Esaki~\cite{ref:tsu-esaki, ref:gehring}, and (3) the WKB approximation to calculate the tunneling probability.

We use these calculated critical current densities to estimate qubit resonance frequencies for FinMETs, and show that qubit frequencies of 4--5~GHz can be obtained with achievable Si barrier thicknesses.
We also fit the critical current results to a model based on the WKB tunneling probability for a rectangular barrier, thus allowing straightforward estimates of the impact of changes in thickness and barrier height on qubit frequencies.


\section{First-principles Approach} \label{sec:approach}

All calculations in this work employ the plane-wave density-functional theory framework with projector-augmented wave (PAW) pseudopotentials~\cite{ref:paw, ref:vasp-paw} as implemented in the Vienna Ab initio Simulation Package (VASP)~\cite{ref:vasp, ref:vasp2} with a plane-wave cutoff of 500~eV. 
Bulk calculations aimed at obtaining accurate electronic energy levels use the hybrid functional of Heyd, Scuseria, and Ernzerhof (HSE)~\cite{ref:HSE03, ref:HSE06}, while calculations for interfacial structure and potential alignment use the PBE functional~\cite{ref:pbe}; tests of the accuracy are described below.
Structural optimizations are performed until the forces are less than 0.01~eV/\AA.

\subsection{Schottky barrier heights} \label{subsec:approach-sbh}

\begin{figure}[h!]
    \centering
    \includegraphics[width=0.9\linewidth]{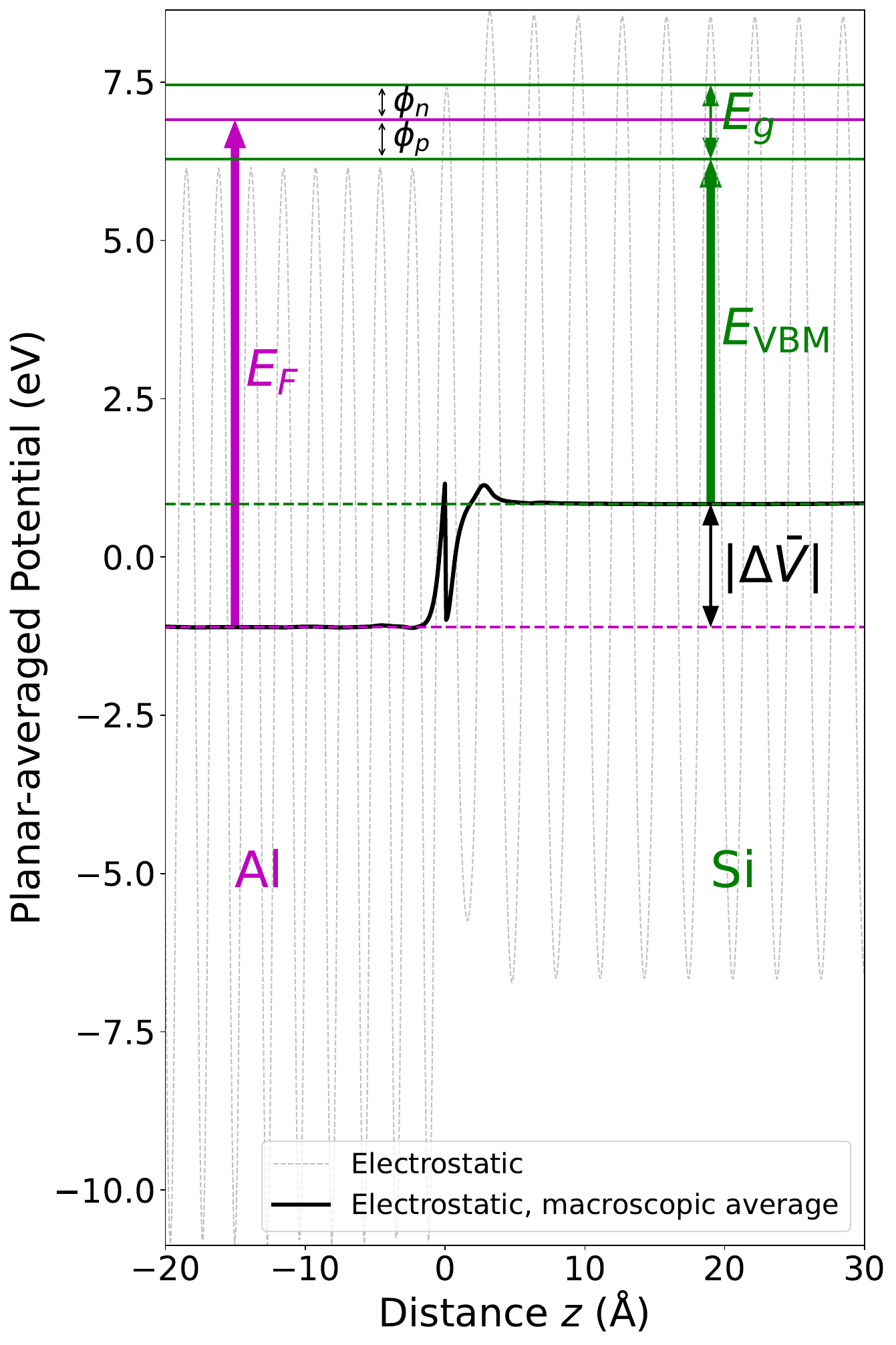}

    \caption{
        Derivation of SBH values at the Al(111)/Si(111) interface with type-A 
        orientation. 
        Shown are the planar-averaged electrostatic potential as a function of distance $z$ along the [111] direction (dashed curve), 
        the macroscopic average of the electrostatic potential (solid curve), 
        the resulting average potential difference between Al and Si $\Delta \bar{V}$, 
        the Fermi level $E_F$ of Al, 
        the valence-band maximum $E_\mathrm{VBM}$ and band gap $E_g$ of Si, 
        and the resulting Schottky barrier heights $\phi_p$ and $\phi_n$.
    }

    \label{fig:schottky}
\end{figure}

To calculate the SBHs, we use the potential alignment method~\cite{ref:vandewalle87,ref:delaney, ref:weston}:
\begin{align}
    \phi_p &= \Delta \bar{V} + E_F - E_\mathrm{VBM}, \label{eqn:psbh} \\
    \phi_n &= E_\mathrm{g} - \phi_p, \label{eqn:nsbh}
\end{align}
where $\phi_p$ and $\phi_n$ are the $p$- and $n$-type SBH, respectively.
All quantities are illustrated in Fig.~\ref{fig:schottky}.
$\Delta \bar{V}$ is the average electrostatic potential difference across the interface (positive if the metal has higher average potential), 
$E_F$ is the metal Fermi level referenced to the average electrostatic potential of the metal, 
$E_\mathrm{VBM}$ is the valence-band maximum (VBM) referenced to the averaged electrostatic potential of the semiconductor, and 
$E_g$ is the band gap. 
Note that the quantities labeled as ``potentials'' are actually potential energies for electrons (in units of eV), as is conventional in band diagrams.
$E_F$, $E_\mathrm{VBM}$, and $E_g$ are obtained from bulk calculations, and $\Delta \bar{V}$ from a calculation of the interface. 

To calculate $\Delta \bar{V}$, we use the macroscopic averaging method~\cite{ref:macroAvg}. First, the electrostatic potential of the interface is averaged over the interface plane, yielding the dashed curve in Fig.~\ref{fig:schottky}. Then, the planar-averaged electrostatic potential $\langle V \rangle_{xy}$ is averaged along the perpendicular direction $z$ according to 
\begin{align}
    \bar{V}(z) = \int_{z-L/2}^{z+L/2} \langle V \rangle_{xy}(z')~dz', \label{eqn:macoAvg}
\end{align}
where $L$ equals the oscillation period along $z$ at the center of the metal (semiconductor) slab for $z$ inside the metal (semiconductor). The resulting $\bar{V}(z)$ is illustrated by the solid curve in Fig.~\ref{fig:schottky}.

We regard the potential alignment method for calculating the SBH as more accurate than the layer-projected density-of-states approach \cite{ref:zhao,ref:gao,ref:wasey}.  In the latter, the determination of the VBM and the conduction-band minimum (CBM) is prone to errors due to the low densities of states near the band edges. 
A key advantage of potential alignment is that the electrostatic potential converges to its bulk value within about two atomic layers from the interface, confirming that the SBH is truly a property of the interface,
and can be used as a boundary condition in calculations that would include, e.g., band bending on longer length scales. 
Similar comments about the potential alignment method being more accurate than the layer-projected density-of-states approach were included in Ref.~\onlinecite{ref:ma}.

\subsection{Bulk calculations} \label{ssubsec:approach-bulk}

All bulk calculations are performed with the HSE hybrid functional~\cite{ref:HSE03, ref:HSE06}. 
For Si we use a Brillouin-zone sampling mesh of $15 \times 15 \times 15$.
We adjust the HSE mixing parameter to reproduce the silicon experimental gap at zero temperature ($E_g$ = 1.17~eV~\cite{ref:si_exp_gap}), yielding a value of 0.256. The corresponding lattice parameter is 5.432~\AA, 
in agreement with the experimental value of 5.431~\AA~\cite{ref:si_exp-lattConst}. 
The resulting $E_\mathrm{VBM}$ value (shown in Fig.~\ref{fig:schottky})
is 5.45~eV, which we use in all our subsequent calculations of SBHs.
We note that this quantity, taken in isolation, is not physically meaningful, since it depends on the specific pseudopotentials used in the calculations.

We optimize the structures of bulk Al and CoSi$_2$ with the same HSE mixing parameter, a Brillouin-zone sampling mesh of $16 \times 16 \times 16$, and the second-order Methfessel-Paxton scheme with a smearing width of 0.2~eV to aid numerical convergence. 
For Al, which has the face-centered cubic (fcc) structure, we calculate an equilibrium lattice parameter of 4.024~{\AA}, 
0.6\% smaller than the experimental value of 4.049~\AA~\cite{ref:al-exp-latt-const}. 
CoSi$_2$ has the CaF$_2$ structure~\cite{ref:stadler} (cubic Fm$\bar{3}$m space group), which can be regarded as a zinc-blende structure with additional Si atoms on all tetrahedral interstitial sites, resulting in a fourfold coordination for Si and an eightfold coordination for Co. The calculated equilibrium lattice parameter is 5.293~{\AA}, 
1.3\% smaller than the experimental value of 5.365~\AA~\cite{ref:cosi2-exp-latt-const}. 

In the actual interfaces, because Si(111) is the substrate, Al and CoSi$_2$ are strained to match silicon as discussed in Sections \ref{subsec:results-al_struct-en} and \ref{subsec:results-cosi2_struct-en}.   
To facilitate application of this strain we describe Al and CoSi$_2$ in a hexagonal unit cell in which the $c$ axis is oriented along the [111] direction of the interface. 
This unit cell contains 3 atoms for Al, and 3 Co atoms and 6 Si atoms for CoSi$_2$. 
Then, fixing the in-plane lattice parameters to their strained values, 
we optimize the perpendicular lattice parameter of each metal.  When we do this fully consistently within HSE, we obtain the the following Fermi levels $E_F$ (referenced to the averaged electrostatic potential) of the resulting optimized structures: 8.02~eV for Al and 9.69~eV for CoSi$_2$.
These values are used in conjunction with our calculations of SBHs based on full HSE interface calculations. 
Because of the high computational cost of HSE calculations for interfaces, we also perform interface calculations using the PBE functional. 
To consistently calculate the SBHs for such interfaces, we use HSE to optimize the perpendicular lattice parameter and calculate the Fermi levels $E_F$ of Al and CoSi$_2$ with in-plane strain values corresponding to those calculated for PBE lattice constants. For this case we obtain 7.98~eV for Al and 9.81~eV for CoSi$_2$.

\subsection{Interface calculations} \label{subsec:approach-interf}

To obtain the average potential difference $\Delta \bar{V}$, we calculate the electrostatic potentials in explicit interface calculations for Al(111)/Si(111) and CoSi$_2$(111)/Si(111). 
Periodic boundary conditions require such calculations to be performed for a superlattice; if the layers are sufficiently thick, properties of a single interface can be obtained.
In addition, the Al/Si interface requires use of large in-plane unit cells due to the lattice mismatch between Al and Si.
The resulting large supercells render the calculations extremely expensive when performed with a hybrid functional.
To make the calculations tractable we use the PBE functional~\cite{ref:pbe} for all our Al/Si interface calculations. 
As shown in \cite{ref:weston}, the error in $\Delta \bar{V}$ from using the PBE functional as opposed to the HSE functional is less than $\sim$0.05~eV. 
Our own test calculations comparing $\Delta \bar{V}$ from PBE and HSE calculations produced a similar result, an uncertainty of $\sim$0.06~eV. 
For CoSi$_2$(111)/Si(111), we do not need to consider large in-plane unit cells because CoSi$_2$(111) is nearly lattice-matched to Si(111)), and hence the superlattices used for interface calculations contain significantly fewer atoms compared to Al(111)/Si(111).  The resulting lower computational expense allowed us to also perform full HSE calculations for a few of the lowest-energy structures of CoSi$_2$(111)/Si(111).

We first repeated the structural optimization procedures for bulk Si, Al, and CoSi$_2$ described in Sec.~\ref{ssubsec:approach-bulk} using the PBE functional, and then used these optimized bulk structures to construct  Al(111)/Si(111) and CoSi$_2$(111)/Si(111) interfaces using a hexagonal-symmetry supercell with the perpendicular lattice parameter along the [111] direction. 
Due to periodic boundary conditions, each supercell contains two metal/semiconductor interfaces. 
Fortunately the symmetry of the underlying materials allows us to choose the numbers of layers in such a way that the superlattice has inversion symmetry, and therefore the two interfaces in each  supercell have the same properties and identical Schottky barrier heights. 

We optimized the interfacial structure by allowing a few layers near each interface (two Al layers, two Si-Si bilayers, and two CoSi$_2$ trilayers) to relax their atomic positions and their interlayer spacings, while also allowing the perpendicular lattice parameter of the supercell to relax.
The in-plane lattice parameters, as well as the interlayer spacings closer to the centers of the Si and metal layers in the supercell were kept fixed. 
Using the resulting optimized structure, we calculate the average electrostatic potential, and subsequently $\Delta \bar{V}$, as described in Refs.~\onlinecite{ref:vandewalle87}, \onlinecite{ref:delaney} and \onlinecite{ref:weston}.
For all results presented in this work, the layer thicknesses and
the number of relaxed layers are sufficient to converge $\Delta \bar{V}$ within 0.015~eV, as documented in the Supplemental Material~\cite{ref:supplemental}. 
Differences in SBHs between different structures are expected to be even more accurate.

For Al(111)/Si(111), we choose a supercell containing 13 Al layers and 12 Si-Si bilayers.
This is based on the analysis of the dependence of $\Delta \bar{V}$ on layer thickness presented in Table~\ref{tab:deltaV_al} in the Supplemental Material~\cite{ref:supplemental}: we see that the $\Delta \bar{V}$ for this supercell is within 0.015~eV from the $\Delta \bar{V}$ for a thicker supercell containing 16 Al layers and 15 Si-Si bilayers. 
Similarly, for each CoSi$_2$(111)/Si(111) structure, we choose a supercell size for which the $\Delta \bar{V}$ is within 0.015~eV from the $\Delta \bar{V}$ for a thicker supercell we considered, as shown in Table~\ref{tab:deltaV_cosi2}. 
The chosen supercell sizes depend on the structure and are shown in Table~\ref{tab:chosenLayerThicknesses}. We make sure that our thinnest CoSi$_2$ slab (10 CoSi$_2$ trilayers) is still thicker than the thickest CoSi$_2$ slab of Ref.~\cite{ref:stadler} (6 CoSi$_2$ trilayers) to avoid the uncertainty 
reported in \cite{ref:stadler}.

\subsection{Interface formation energies} \label{subsec:form-en}

The interface formation energy can be expressed as~\cite{ref:hamann, ref:delaney}
\begin{align} 
    \gamma^\text{f} 
        &= \frac{1}{2A} \left[ E^\text{tot} - N_\mathrm{Si} E^\text{tot}_\text{Si} - N_\mathrm{metal} E^\text{tot}_\text{metal} \right], \label{eqn:formationEnergy}
\end{align}
where $\gamma^\text{f}$ is the formation energy of a single interface per
unit area, $A$ is the in-plane area of the supercell, $E^\text{tot}$ is
the total energy of the supercell (which contains two interfaces),
$N_\mathrm{Si}$ is the number of Si atoms in the Si layers, 
and $E^\text{tot}_\text{Si}$ is the total energy per Si atom in bulk Si. 
For Al(111)/Si(111), $N_\mathrm{metal}$ is the number of Al atoms in the supercell, and $E^\text{tot}_\text{metal}$ is the total energy per Al atom in bulk \emph{strained} Al. 
For CoSi$_2$(111)/Si(111), $N_\mathrm{M}$ is the number of CoSi$_2$ units in the CoSi$_2$ layers, and $E^\text{tot}_\text{metal}$ is the total energy per CoSi$_2$ unit in bulk strained CoSi$_2$. 
Choosing strained metal as the reference ensures that there is no dependence on the thickness of the metal layers.

\subsection{Josephson critical current estimates} \label{subsec:tunneling}

In this section we describe our methodology to derive Josephson critical current densities.   
We use a number of approximations, justified by the fact that our goal is not to obtain values with the highest accuracy but rather to examine trends.

To estimate the critical current density $J_c$ of a Josephson junction, we use the formulation of $J_c$ by Ambegaokar and Baratoff~\cite{ref:ambegaokar,ref:tinkham}, 
\begin{align}
    J_c(s,T) = \frac{\pi/e}{R_\mathrm{n}(s,T) A_\mathrm{eff}}  \frac{\Delta(T)}{2} \tanh \left( \frac{\Delta(T)}{2 k_B T} \right), \label{eqn:jc}
\end{align}
where $s$ is the thickness of the tunnel barrier, $T$ is the temperature, $e$ is the elementary charge, $R_\mathrm{n}$ is the normal-state (non-superconducting) resistance of the junction, 
$A_\mathrm{eff}$ is the effective cross-sectional area of the junction, $\Delta$ is the superconducting gap parameter (which depends on temperature), and $k_B$ is the Boltzmann constant. 

From Ohm's Law, 
\begin{align}
    R_\mathrm{n}(s,T) A_\mathrm{eff} = \frac{\mathcal{V}}{J(s,\mathcal{V},T)}, \label{eqn:RA}
\end{align}
where $\mathcal{V}$ is the magnitude of the applied voltage across the junction, and $J$ is the resulting normal-state current density. 
We use 
the tunneling model developed by both Simmons~\cite{ref:simmons1963, ref:simmons1964} and Tsu and Esaki~\cite{ref:tsu-esaki, ref:gehring}  
to calculate $J_e$, the contribution to $J(s,\mathcal{V},T)$ due to tunneling of electrons through the barrier below the CBM: 
\begin{align}
    J_e(s,\mathcal{V},T) = \frac{4\pi e}{h^3} m^*_\mathrm{M} \int_{0}^{E_z^\mathrm{max}} \mathcal{T}_e(s,\mathcal{V},E_z) ~\mathcal{N}_e(T,\mathcal{V},E_z) ~dE_z , \label{eqn:jvt}
\end{align}
where 
$E_z$ is the longitudinal energy along the tunneling direction $z$, 
$h$ is Planck's constant, $\mathcal{T}_e$ is the electron transmission coefficient through the barrier, and $\mathcal{N}_e$ is the electron ``supply function'' of the junction, a function indicating how many electrons participate in the tunneling, which we describe in more detail later below. 
The electrons in the metal electrodes are assumed to have parabolic dispersion and a density of states corresponding to a free-electron gas with an effective mass $m^*_\mathrm{M}$. 
The integration over longitudinal energy $E_z$ is carried out over the relevant range 
$E_z < E_z^\mathrm{max}$, 
as detailed below. 

\begin{figure}[h!]
    \centering
    \includegraphics[width=\linewidth]{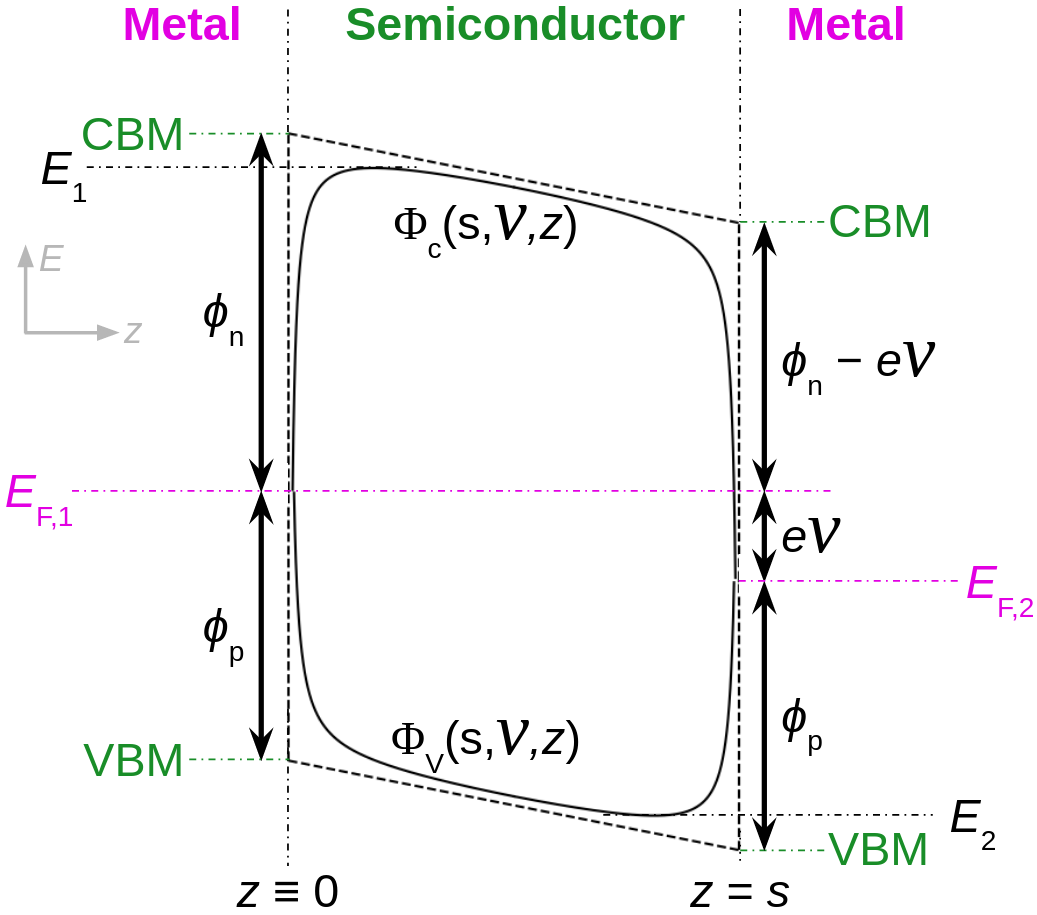}

    \caption{
        Potential energy landscape 
        across a metal/semiconductor/metal junction where the right metal electrode is at a voltage $\mathcal{V}$ higher than the left metal electrode. Shown are the metal Fermi levels ($E_\mathrm{F,1}$, $E_\mathrm{F,2}$), positions of CBM and VBM at the metal contacts, $n$- and $p$-type SBHs ($\phi_n$, $\phi_p$), and semiconductor thickness $s$. 
        The variation of the CBM $\Phi_\mathrm{C}$ and VBM $\Phi_\mathrm{V}$ across the junction 
        is shown for the case with (solid curves) and without (dashed lines) image charge correction. 
        The maximum of $\Phi_\mathrm{C}$, $E_1$, and the minimum of $\Phi_\mathrm{V}$, $E_2$, are also indicated.
    }

    \label{fig:tunneling}
\end{figure}

When a voltage $\mathcal{V}$ is applied to the junction, 
the potential energy across the barrier will vary as a function of the distance $z$, as illustrated in Fig.~\ref{fig:tunneling}. The right electrode is at a higher voltage than the left, and therefore the electrons will tunnel through the barrier from left to right. 
The dashed lines in Fig.~\ref{fig:tunneling} show the variation of the 
CBM ($\Phi_\mathrm{C}$) and VBM ($\Phi_\mathrm{V}$) 
in the absence of an image charge correction, and the solid lines include this correction, as described below.
The maximum longitudinal energy $E_z^\mathrm{max}$ of the tunneling electrons equals $E_1$, the maximum of $\Phi_\mathrm{C}$, which corresponds to the top of the barrier.

The potential energies $\Phi_\mathrm{C}(s,\mathcal{V},z)$ and $\Phi_\mathrm{V}(s,\mathcal{V},z)$ are given by 
\begin{align}
    \Phi_\mathrm{C}(s,\mathcal{V},z) &= E_\mathrm{F,1} + \phi_n - (e\mathcal{V}/s)z + E_\mathrm{imag}(s,z), \label{eqn:PhiC} \\
    \Phi_\mathrm{V}(s,\mathcal{V},z) &= E_\mathrm{F,1} - \phi_p - (e\mathcal{V}/s)z - E_\mathrm{imag}(s,z), \label{eqn:PhiV}
\end{align}
where $E_\mathrm{F,1}$ is the Fermi level of the left metal electrode,
$\phi_n$ is the $n$-type SBH, and $E_\mathrm{imag}$ is the image charge correction term for metal/semiconductor/metal junctions~\cite{ref:imag_corr,ref:gehring}:
\begin{align}
    E_\mathrm{imag}(s,z) = \frac{e^2}{16\pi \epsilon_r \epsilon_0}  \sum_{j=0}^{\infty} 
    \bigg[ 
        - \frac{1}{js+z} &- \frac{1}{(j+1)s-z}
    \nonumber \\ 
        &+ \frac{2}{(j+1)s} 
    \bigg]. \label{eqn:imgChg}
\end{align}
Here, $\epsilon_r$ is the dielectric constant of the semiconductor (11.7 for Si~\cite{ref:Si_dielConst}) and $\epsilon_0$ is the vacuum permittivity. 
Without the image charge correction, both $\Phi_\mathrm{C}(s,\mathcal{V},z)$ and $\Phi_\mathrm{V}(s,\mathcal{V},z)$ become triangular barriers shown by the dashed lines in Fig.~\ref{fig:tunneling}.

To evaluate the integrand in Eq.~\eqref{eqn:jvt}, 
we opt to use the WKB approximation to calculate $\mathcal{T}_e$ and 
use the Fermi-Dirac distributions of the electrons in the metal electrodes to calculate $\mathcal{N}_e$~\cite{ref:gehring}:
\begin{align}
    \mathcal{T}_e(s,\mathcal{V},E_z) &= \exp \left[ -\frac{2}{\hbar} \int_{z_{1\mathrm{C}}}^{z_{2\mathrm{C}}} \sqrt{2m^*_e (\Phi_\mathrm{C}(s,\mathcal{V},z) - E_z)} dz \right] \label{eqn:trans-coeff} \\
    \mathcal{N}_e(T,\mathcal{V},E_z) &= k_B T \ln \left\{ 
        \frac{
            1 + \exp [ -(E_z-E_\mathrm{F,1})/(k_B T) ]
        }{ 
            1 + \exp [ -(E_z-E_\mathrm{F,2})/(k_B T) ]
        } 
    \right\}, \label{eqn:supply-func}
\end{align}
where 
$E_\mathrm{F,2} = E_\mathrm{F,1} - e\mathcal{V}$ is the Fermi level of the right metal electrode, 
$m^*_e$ is the tunneling effective mass for electrons (derived from the complex band structure as described below), 
$\hbar$ is the reduced Planck's constant, 
and $z_{1\mathrm{C}}$ and $z_{2\mathrm{C}}$ are the positions where $\Phi_\mathrm{C}(s,\mathcal{V},z) = E_z$. 

Besides the contribution from electrons tunneling through 
the barrier below 
the CBM described above, there is also a contribution from holes tunneling through 
the barrier above 
the VBM from the right to the left electrodes. Following the derivations in Ref.~\onlinecite{ref:gehring}, we can derive the analogs of Eqs. \eqref{eqn:jvt}, \eqref{eqn:trans-coeff}, and \eqref{eqn:supply-func} for holes:
\begin{align}
    J_h(s,\mathcal{V},T) &= \frac{4\pi e}{h^3} m^*_\mathrm{M} \int_{E_2}^{\infty} \mathcal{T}_h(s,\mathcal{V},E_z) ~\mathcal{N}_h(T,\mathcal{V},E_z) ~dE_z, \label{eqn:jvt_h} \\
    \mathcal{T}_h(s,\mathcal{V},E_z) &= \exp \left[ -\frac{2}{\hbar} \int_{z_{1\mathrm{V}}}^{z_{2\mathrm{V}}} \sqrt{2m^*_h (E_z - \Phi_\mathrm{V}(s,\mathcal{V},z))} dz \right] \label{eqn:trans-coeff_h} \\
    \mathcal{N}_h(T,\mathcal{V},E_z) &= k_B T \ln \left\{ 
        \frac{
            1 + \exp [ +(E_z-E_\mathrm{F,2})/(k_B T) ]
        }{ 
            1 + \exp [ +(E_z-E_\mathrm{F,1})/(k_B T) ]
        } 
    \right\}, \label{eqn:supply-func_h}
\end{align}
where $E_2$ is the minimum of $\Phi_\mathrm{V}$ as shown in Fig.~\ref{fig:tunneling}, 
$m^*_h$ is the tunneling effective mass for holes, 
and $z_{1\mathrm{V}}$ and $z_{2\mathrm{V}}$ are the positions where $\Phi_\mathrm{V}(s,\mathcal{V},z) = E_z$.
The total normal-state current density $J(s,\mathcal{V},T)$ that enters Eq.~\eqref{eqn:RA} is then
\begin{align}
    J(s,\mathcal{V},T) &= J_e(s,\mathcal{V},T) + J_h(s,\mathcal{V},T). \label{eqn:Jtot}
\end{align}

In the present work, 
we calculate the critical current densities $J_c$ for various values of $s$.  
We use $T = 0.02$~K, a typical operating temperature of superconducting qubits~\cite{ref:krantz}, which is significantly smaller than the superconducting transition temperatures $T_c$ of both Al (1.2~K~\cite{ref:al-Tc}) and CoSi$_2$ (1.26~K~\cite{ref:cosi2-superconducts}). 
According to the BCS theory, which describes Al and CoSi$_2$ relatively well~\cite{ref:al-Tc, ref:cosi2-superconducts}, the gap parameter $\Delta(T)$ is practically unchanged for $T < 0.5 T_c$~\cite{ref:tinkham_bcs}, and therefore we treat $\Delta$ as constant at these temperatures. We choose $\Delta$ equal to 0.2~meV for Al (which is the value reported in Ref.~\onlinecite{ref:supgap_al} from extrapolating the experimental data to 0~K) and 0.189~meV for CoSi$_2$ (from averaging the measured gap parameters reported in Ref.~\onlinecite{ref:supgap_cosi2} at 0.37~K [$0.3T_c$ of CoSi$_2$]).
Note that we evaluate the normal-state current density $J(s,\mathcal{V},T)$ at the operating temperature $T$ of the superconducting junction.
We have checked that the calculated $J_c$ values vary weakly with $T$ for $T < 0.5T_c$.

We take the electron effective mass $m^*_\mathrm{M}$ in the metal electrodes to be $\sim$1.4$m_0$ for Al~\cite{ref:al-mEff} and $\sim$$m_0$ for CoSi$_2$~\cite{ref:cosi2-mEff}, where $m_0$ is the free-electron mass. 
To obtain the tunneling effective masses for electrons ($m^*_e$) and holes ($m^*_h$), we take the complex band structure of Si[111] computed in Ref.~\onlinecite{ref:laux} and fit the imaginary part to $\sqrt{2 m^*_e (E_\mathrm{CBM}-E)}/\hbar$ and $\sqrt{2 m^*_h (E-E_\mathrm{VBM})}/\hbar$, respectively. We obtain $m^*_e = 0.19 m_0$ and $m^*_h = 0.08 m_0$.
For the image charge correction, we carry out the sum over $j$ in Eq.~\eqref{eqn:imgChg} until $j=11$ (as in Ref.~\onlinecite{ref:gehring}), up to which we find that the sum has converged to within 3\% of the sum up to $j=10001$. 
Finally, we use sufficiently small values of $\mathcal{V}$ such that the junction is in the Ohmic region~\cite{ref:dorneles} and therefore Eq.~\eqref{eqn:RA} holds, yielding $R_\mathrm{n}(s,T) A_\mathrm{eff}$ values that vary weakly with $\mathcal{V}$. 

To allow comparison between our estimated $J_c$ values and the desired values for actual silicon-based METs, we estimate the qubit resonant frequencies based on our $J_c$ values. The resonant frequency is given by~\cite{ref:krantz}
\begin{align}
    f_q = (1/h) \left[ \sqrt{8 E_J E_C} - E_C \right] \approx (1/h)\sqrt{8 E_J E_C}, \label{eqn:qubit-freq}
\end{align}
where $E_J$ is the energy stored in the Josephson junction, 
and $E_C$ is the energy stored in the shunt capacitor. 
The approximation in the above equation is valid for transmons, for which typically $E_J \geq 50 E_C$~\cite{ref:krantz}. 
The energies $E_J$ and $E_C$ are given by~\cite{ref:krantz}
\begin{align}
    E_J &= \frac{\hbar}{2e} J_c A_\mathrm{eff}, \label{eqn:EJ}
    \\
    E_C &= \frac{e^2}{2 C} = \frac{e^2 s}{2 \epsilon_r \epsilon_0 A_\mathrm{eff}}, \label{eqn:EC}
\end{align}
where we have used the parallel-plate capacitor formula for the capacitance $C$. 
Plugging in these equations into Eq.~\eqref{eqn:qubit-freq} yields
\begin{align}
    f_q \approx \sqrt{ \frac{e}{\pi h \epsilon_0} \frac{J_c s}{\epsilon_r} }, \label{eqn:qubit-freq-Jcs}
\end{align}
and thus, approximately, $f_q$ only depends on the semiconductor's dielectric constant $\epsilon_r$ and thickness $s$, and the Josephson critical current density $J_c$.

Novel silicon-based METs, e.g. FinMETs~\cite{ref:finmet}, aim to achieve a qubit resonant frequency in the range 4--5~GHz, the same range as the predecessor METs made of Al/AlO$_x$ junctions~\cite{ref:met_alox}. Therefore we will use Eq.~\eqref{eqn:qubit-freq-Jcs} in combination with our estimated values for $J_c$ as a function of $s$ to estimate the resulting $f_q$'s and compare them with the desired range of 4--5~GHz.


\section{Results: \texorpdfstring{A\MakeLowercase{l}(111)/S\MakeLowercase{i}(111)}{Al(111)/Si(111)}} \label{sec:results-al}


\subsection{Structures and formation energies} \label{subsec:results-al_struct-en}

\begin{figure}[h!]
    \centering
    \includegraphics[width=0.95\linewidth]{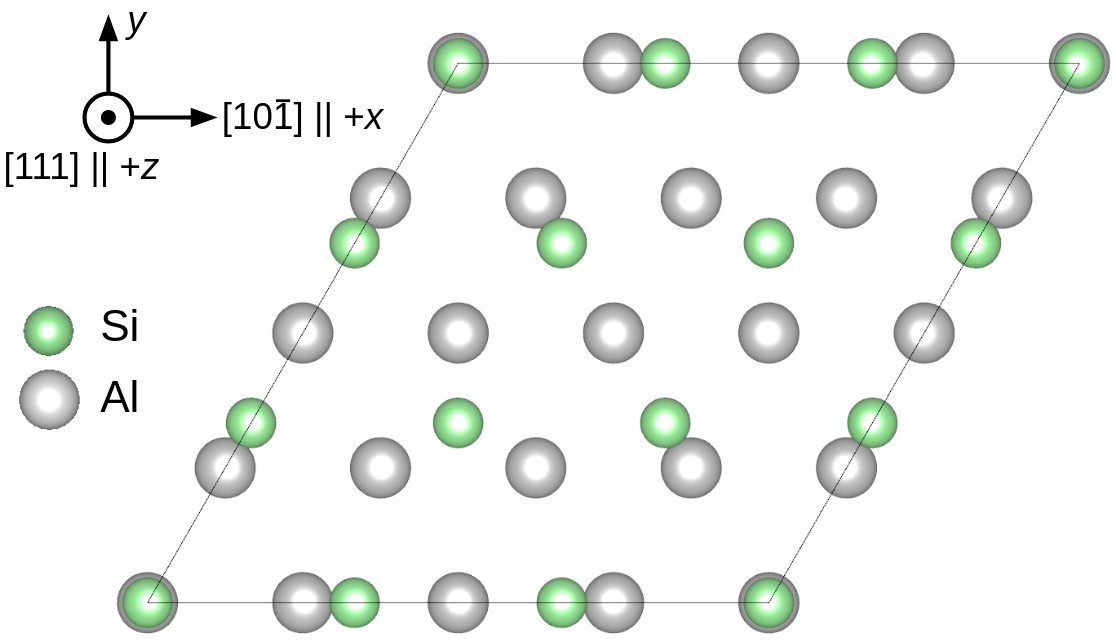}

    \caption{
        In-plane unit cell of the Al(111)/Si(111) hexagonal supercell, showing the positions of Al and Si atoms near the interface, with three repeat units of Si matching 4 repeat units of Al. 
    }

    \label{fig:alsi_xy}
\end{figure}

Forcing the in-plane periodicity of Al to match that of Si in a (1$\times$1) interfacial unit cell would lead to a huge in-plane strain, since the lattice parameters of Al and Si differ by 35\%. 
A commensurate interface that minimizes the strain can be identified by matching four in-plane lattice parameters of Al to three in-plane lattice parameters of Si, leading to a remaining strain of only $\sim$1\% in Al. 
The resulting structure, which was experimentally observed ~\cite{ref:legoues, ref:yapsir, ref:fortuin}, is illustrated in Fig.~\ref{fig:alsi_xy}. 
Due to the large lattice parameters of the supercell, we find that a Brillouin-zone sampling mesh of $4 \times 4 \times 1$ is sufficient to converge $\Delta \bar{V}$ to within 0.01~eV.

\begin{figure}[h!]
    \centering
    \includegraphics[width=\linewidth]{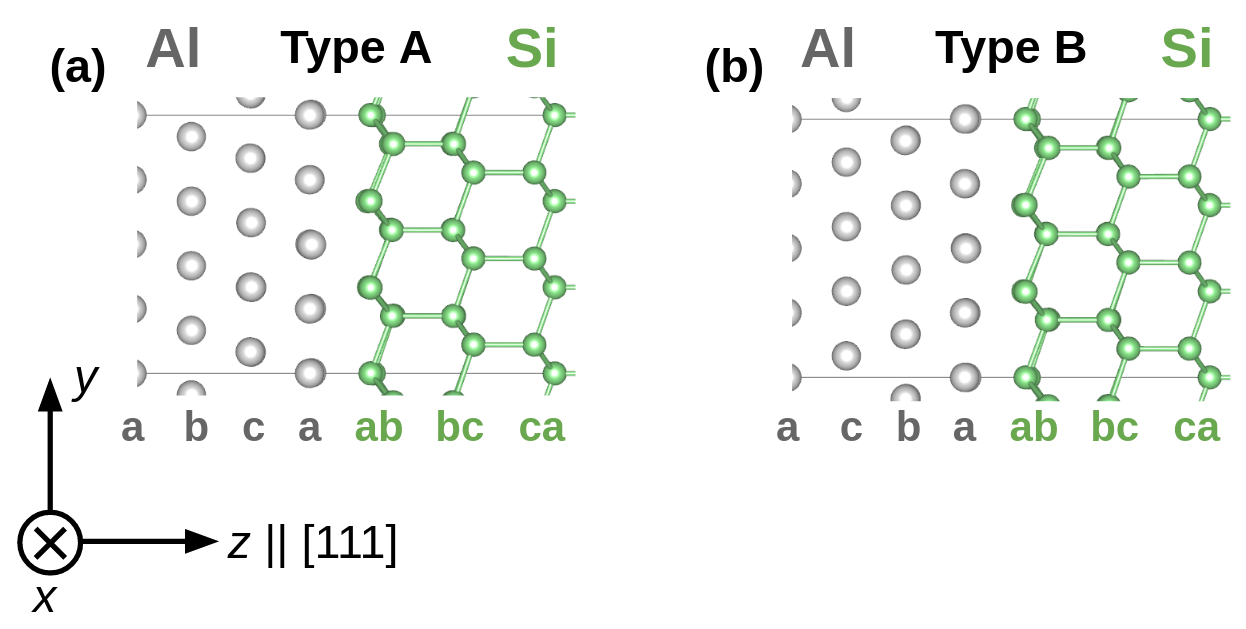}

    \caption{
        Side view of Al(111)/Si(111) structures near the interface, shown here for Al and Si interfacial layers being both ``a'' layers in the fcc stacking.
        (a) Type-A orientation, where Al and Si have the same fcc stacking sequence (abc along the [111] direction). 
        (b) Type-B orientation, where Al and Si have opposite fcc stacking sequences (acb for Al and abc for Si).
    }

    \label{fig:alsi_long}
\end{figure}

Figure~\ref{fig:alsi_long} shows a side view of the Al(111)/Si(111) structure near the interface. Two possibilities exist for the relative orientation of the Si and Al layers~\cite{ref:miura_typeB}: type A, where the fcc stacking sequence is the same for both the metal and the semiconductor, and type B, where the fcc stacking sequences are different. 
Note that both Figures \ref{fig:alsi_xy} and \ref{fig:alsi_long} show a particular case where on each Al layer, there is one Al atom at the same in-plane position as a Si atom at the interface. 
In principle, other structures can occur where none of the Al atoms have the same in-plane position as the interfacial Si atoms. 
In practice, we found that these other structures all have very similar energies. 
For each orientation type, starting from the structures shown in the figures, when we allowed all Al atoms (including the central layers) to relax in the (111) plane (in addition to allowing the layers near the interface to relax in all directions), 
the Al atoms in the central layers practically did not move, 
and the resulting relaxed structure yielded a total energy within 0.3~meV per ($1\times 1$)Si and a $\Delta \bar{V}$ within 5~meV from the results corresponding to fixing the central Al atoms. 
We also have relaxed the structures in which we uniformly displace the Al atoms along the (111) plane such that one Al atom at each layer is now at the same in-plane position as a Si atom at the \emph{second} layer from the interface. 
We found that the central Al atoms also practically did not move, and that the energies and $\Delta \bar{V}$ are within 0.1~meV per ($1\times 1$)Si and 2~meV from the undisplaced structure for type B, and are within 0.6~meV per ($1\times 1$)Si and 2~meV for type A. 
These suggest that the particular in-plane positions of the Al atoms relative to the Si atoms shown in Figures \ref{fig:alsi_xy} and \ref{fig:alsi_long} are energetically as favorable as other relative in-plane positions, and that all of these positions have very similar SBHs.
Additionally, we examined 24 other relative in-plane positions and calculated their total energies without relaxation, and found that the energies were all within 0.3~meV per ($1\times 1$)Si from each other, again suggesting that there is no energetic preference for a particular in-plane position of Al relative to Si. 

We also considered another possible case where the ``a'' layer of Si at the interface is removed. For both types A and B corresponding to this case, the energies per ($1\times 1$)Si after relaxation are $\sim$6~eV higher, demonstrating that these structures are far less energetically favorable.
Due to all of the above reasons, in the present work we focus on the particular structures shown in Figures \ref{fig:alsi_xy} and \ref{fig:alsi_long}, and proceed with relaxing a few layers near the interface as described next.

For both orientation types, we relax two Al layers and two Si-Si bilayers near each interface. 
Relaxing three Al layers and three Si-Si bilayers changes the $\Delta \bar{V}$ value by merely $\sim$1~meV, as seen in Table~\ref{tab:deltaV_al} in the Supplemental Material~\cite{ref:supplemental}. 
The magnitudes of the relaxations of the atoms in these layers are reported in Table~\ref{tab:alsi_rlx}. Since the atoms within a particular layer may have different magnitudes of relaxation, the table reports the range of such displacements. 
We see that the largest Al displacement is merely 0.045~{\AA} for type A and 0.029~{\AA} for type B. The Si atoms tend to move more, the largest displacement being 0.182~{\AA} for type A and 0.156~{\AA} for type B. 

\begin{table}[h!]
\caption{
    Magnitudes of relaxations of Al(111)/Si(111) layers near the interface for type-A and type-B orientations. ``1'' indicates the layer at the interface, ``2'' indicates the next layer further away from the interface, and so on. 
    \label{tab:alsi_rlx}
}
\begin{ruledtabular}
\begin{tabular}{lcc}
    Layer & \multicolumn{2}{c}{Range of displacements (\AA)} \\
    & A & B \\
    \hline
    Al $3$ (fixed) & 0 & 0 \\
    Al $2$                & 0.008--0.041 & 0.011--0.027 \\
    Al $1$ (at interface) & 0.022--0.045 & 0.009--0.029 \\
    Si $1$ (at interface) & 0.061--0.182 & 0.063--0.156 \\
    Si $2$                & 0.013--0.057 & 0.026--0.057 \\
    Si $3$                & 0.006--0.035 & 0.012--0.040 \\
    Si $4$                & 0.007--0.010 & 0.008--0.011 \\
    Si $5$ (fixed) & 0 & 0 \\
\end{tabular}
\end{ruledtabular}
\end{table}

Table~\ref{tab:alsi_layerSpacing} shows the interlayer spacings between adjacent layers. The spacings between two layers that are both fixed correspond to the equilibrium bulk values (bulk Si and strained bulk Al). We see that the spacings near the fixed layers are very close to the  bulk spacings. 
Moreover, the Al-Si interlayer distance across the interface is 2.343~{\AA} for type A and 2.353~{\AA} for type B, very close to the average of the interlayer spacing of strained bulk Al, 2.311~{\AA}, and the larger interlayer spacing of bulk Si, 2.368~{\AA} (which equals the Si-Si bond length in the bulk as calculated with PBE).

\begin{table}[h!]
\caption{
    Interlayer spacings of Al(111)/Si(111) near the interface.
    \label{tab:alsi_layerSpacing}
}
\begin{ruledtabular}
\begin{tabular}{lcc}
    Layers & \multicolumn{2}{c}{Interlayer spacing (\AA)} \\
                                     & A     & B \\
    \hline
    Al 4 $-$ Al 3 (both fixed)       & 2.311 & 2.311 \\
    Al 3 (fixed) $-$ Al 2            & 2.313 & 2.314 \\
    Al 2 $-$ Al 1                    & 2.304 & 2.305 \\
    Al 1 $-$ Si 1 (across interface) & 2.343 & 2.353 \\
    Si 1 $-$ Si 2                    & 0.849 & 0.848 \\
    Si 2 $-$ Si 3                    & 2.377 & 2.377 \\
    Si 3 $-$ Si 4                    & 0.785 & 0.785 \\
    Si 4 $-$ Si 5 (fixed)            & 2.370 & 2.370 \\
    Si 5 $-$ Si 6 (both fixed)       & 0.789 & 0.789 \\
    Si 6 $-$ Si 7 (both fixed)       & 2.368 & 2.368 \\
\end{tabular}
\end{ruledtabular}
\end{table}

Table \ref{tab:alsi_energy_SBH} shows the interface formation energies per $(1 \times 1)$ Si for all considered Al(111)/Si(111) structures. 
This energy equals the formation energy per unit area $\gamma^\mathrm{f}$, Eq.~\eqref{eqn:formationEnergy}, multiplied by the area $A_{(1 \times 1) \mathrm{Si}} = (\sqrt{3}/4) a_\mathrm{Si}^2=12.9~\mathrm{\AA}^2$. (Here $a_\mathrm{Si}$ is the lattice parameter of Si, which we calculate to be 5.469~{\AA} using PBE.) 
The energies for both orientations are very similar to each other, differing by at most 0.01~eV. 
We attribute this to the weakly covalent character of the bonding across the interface, leading to very similar atomic structures (Fig.~\ref{fig:alsi_long}) and layer-projected densities-of-states (Fig.~\ref{fig:lpdos_alsi} in the Supplemental Material~\cite{ref:supplemental}).

Our calculated energy for the type-A orientation is 0.12~eV higher than the value calculated by molecular dynamics with empirical interatomic potentials~\cite{ref:wu}. We note, however, that Ref.~\onlinecite{ref:wu} did not make it clear what they used as reference energies in calculating the interface energy, e.g., strained bulk Al [as we do in Eq.~(\ref{eqn:formationEnergy})] or unstrained bulk Al, or relaxed/unrelaxed surface slabs of Al and Si.

\begin{table}[h!]
\caption{
    Interface formation energies per (1 $\times$ 1) in-plane Si ($\gamma^\text{f} A_{(1 \times 1) \mathrm{Si}}$),  and \textit{p}- and \textit{n}-type Schottky barrier heights ($\phi_n$, $\phi_p$) of Al(111)/Si(111). 
    \label{tab:alsi_energy_SBH}
}
\begin{ruledtabular}
\begin{tabular}{cccc}
    Orientation type & $\gamma^\text{f} A_{(1 \times 1) \mathrm{Si}}$ (eV) & $\phi_p$ (eV) & $\phi_n$ (eV) \\
    \hline
    A & 0.55 & 0.581 & 0.587 \\
    B & 0.56 & 0.577 & 0.590 \\
\end{tabular}
\end{ruledtabular}
\end{table}

\subsection{Schottky barrier heights} 
\label{subsec:results-al_struct-sbh}

The potential alignment for the interface with type-A orientation is shown in  Fig.~\ref{fig:schottky}.
The macroscopically averaged potential becomes flat within 
about two atomic layers from the interface, 
illustrating convergence to the bulk value. The absence of any slope in the central regions also indicates that the two interfaces in the supercell are truly equivalent.
The kink in the macroscopically averaged potential near the interface is an artifact of the abrupt change in the period used for macroscopic averaging, which is clearly different for Al and Si. 
The difference between the flat regions of the macroscopically averaged potential yields $\Delta \bar{V}=-1.95$~eV.
Per the convention introduced in Sec.~\ref{subsec:approach-interf}
this $\Delta \bar{V}$ value is negative because the macroscopically averaged potential is lower on the metal side.
Combining this $\Delta \bar{V}$ with the bulk values reported in Sec.~\ref{ssubsec:approach-bulk}, we obtain the \textit{p}-type SBH 
$\phi_p=\Delta \bar{V} + E_F - E_\mathrm{VBM}=-1.95+7.98-5.45=0.58$~eV.
Combined with the Si band gap, $E_g=1.17$~eV, this yields an \textit{n}-type Schottky barrier height SBH $\phi_n=0.59$~eV. 

Our calculations for the type-B orientation yield the same $\textit{p}$- and $\textit{n}$-type SBHs to within 0.01~eV, as shown in Table~\ref{tab:alsi_energy_SBH}. 
Again, this is due to the very similar atomic and electronic structures between the two orientations, as discussed in the previous section. 
We estimate an uncertainty of $\sim$0.06~eV from the differences in $\Delta \bar{V}$ observed in our PBE vs.\ HSE test calculations. 
However, we expect the differences between our calculated SBHs to be more accurate, as systematic errors tend to cancel~\cite{ref:cvdw_casi2}.

Experimentally measured $n$-type SBHs are in the range 0.67--0.79~eV~\cite{ref:miura_typeA,ref:miura_typeB}.
A very careful study on epitaxial Al(111)/Si(111) was performed by Miura {\it
et al.}~\cite{ref:miura_typeA}; they verified the 4/3 alignment noted in Fig.~\ref{fig:alsi_xy}, and also checked the orientation of the Al film relative to the Si substrate.  
For a sample with type-A orientation, they obtained $\phi_n$ = 0.68~eV from capacitance-voltage measurements at 200 K.  Values obtained with current-voltage and internal photoemission methods were within 0.01 eV.
They also observed a slight decrease in SBH for this sample as the temperature decreased.
For another sample, in which the Al film was still (111)-oriented but had grains rotating with a random angle distribution in the plane parallel to the interface, they measured $\phi_n$ = 0.77~eV.
For the type-A interface, our calculated value (at 0 K) is 0.59~eV, $\sim$0.1~eV lower than experiment.
The measurements by Miura {\it et al.}~\cite{ref:miura_typeA} seem to indicate that deviations from the pure type-A orientation would lead to an increase in $\phi_n$, which is something we do not observe in our calculations.
The presence of grain boundaries in the experimental samples may affect the measured SBH.

\subsection{Josephson currents} \label{subsec:results-al-current}

Using our calculated SBHs, Table~\ref{tab:alsi_jc} presents estimates of the Josephson critical current densities, Eq.~\eqref{eqn:jc}, for various Si barrier thicknesses in the Josephson junction. 
We include thicknesses ranging from 5~nm to 20~nm, encompassing the thicknesses 5--10~nm for which FinMET devices are expected to have appreciable tunneling currents~\cite{ref:finmet}. 
For each thickness, the critical current densities are practically the same for both type-A and type-B orientations because their SBHs are within 0.01~eV from each other. 
The values in Table~\ref{tab:alsi_jc} include image-charge corrections; without those corrections, the current densities are $\sim$$2\times$ smaller. 

\begin{table}[h!]
\caption{
    Estimates of Josephson critical current densities ($J_c$) and qubit resonant frequencies ($f_q$) for Al(111)/Si(111) junctions, for various Si barrier thicknesses $s$. 
    \label{tab:alsi_jc}
}
\begin{ruledtabular}
\begin{tabular}{rll}
    $s$ (nm) & \multicolumn{1}{c}{$J_c$ (A/m$^2$)}  & \multicolumn{1}{c}{$f_q$ (GHz)} \\
     & A \& B    & A \& B \\
    \hline
     5 & $3 \times 10^{  5}$ & 32 \\
     6 & $2 \times 10^{  4}$ & 10 \\
     7 & $2 \times 10^{  3}$ & 3.5 \\
    10 & $2$ & 0.1 \\
    20 & $10^{-10}$ & $10^{-6}$ \\
\end{tabular}
\end{ruledtabular}
\end{table}

We now compare our estimated critical current densities $J_c$ with the desired values for FinMET devices. Using Eq.~\eqref{eqn:qubit-freq-Jcs}, we estimate the qubit resonant frequencies $f_q$ and tabulate the results in Table~\ref{tab:alsi_jc}. 
We note that $f_q$ increases as thickness decreases, 
and that the desired qubit frequencies of 4--5~GHz~\cite{ref:met_alox} are potentially achievable for thicknesses in the range 6--7~nm.  

\begin{figure}[h!]
    \centering
    \includegraphics[width=\linewidth]{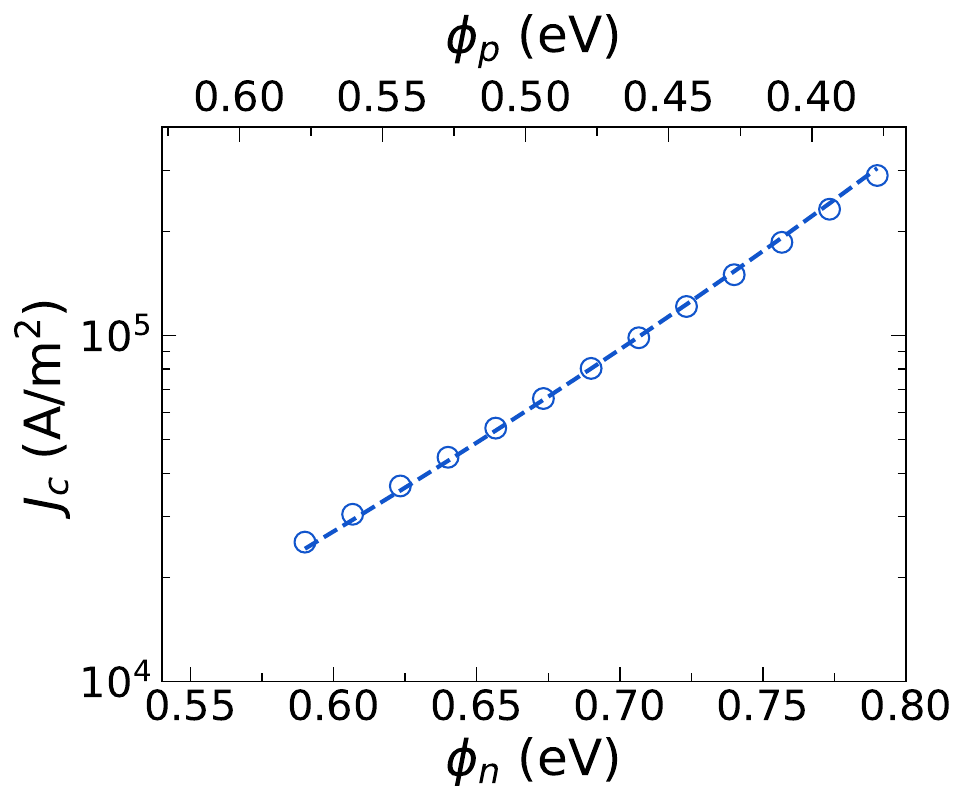}

    \caption{
        Estimated $J_c$ as a function of SBH of Al(111)/Si(111) at Si thickness $s$ = 6~nm (circles) along with the fit to the WKB tunneling probability formula of holes tunneling through a rectangular barrier of height $\phi_p$ (dashed curve).
    }

    \label{fig:alsi_jc}
\end{figure}

It is fruitful to discuss how $J_c$ depends on the SBHs. 
Figure~\ref{fig:alsi_jc} shows the calculated $J_c$ at $n$-SBHs from 0.59~eV to 0.79~eV (corresponding to $\phi_p$ from 0.58~eV to 0.38~eV), a range that includes $n$-SBHs from both our calculations and experimental measurements discussed in the previous section.
For these SBHs, the tunneling processes are actually dominated by holes: the hole contributions $J_h$, Eq.~\eqref{eqn:jvt_h}, are $\sim$100--1000~A/m$^2$, whereas the electron contributions $J_e$, Eq.~\eqref{eqn:jvt}, are merely $\sim$0.01--0.1~A/m$^2$. We attribute this to the tunneling effective mass for holes, $m^*_h = 0.08m_0$, being significantly smaller than the tunneling effective mass for electrons, $m^*_e = 0.19m_0$ (see  Sec.~\ref{subsec:tunneling}). These effective masses appear in the exponents of the transmission coefficients for electrons and holes [Eqs.~\eqref{eqn:trans-coeff} and \eqref{eqn:trans-coeff_h}], and therefore significant differences in these masses lead to differences of several orders of magnitude in $J_h$ and $J_e$. 

Because the hole contributions dominate, the dependence of $J_c$ in Fig.~\ref{fig:alsi_jc} on the SBHs may be approximated by the dependence of the tunneling rate of holes on $\phi_p$. As a simple model we approximate the hole tunneling barrier as a rectangular barrier with uniform height equal to $\phi_p$ across the entire junction, and then use the WKB approximation to calculate the tunneling probability. 
We therefore fit our $J_c$ according to
\begin{align}
    J_c = J_0 \exp{ \left( -\frac{2s}{\hbar} \sqrt{2 m^*_h \phi_p} \right) }, \label{eqn:Jc-fit-psbh}
\end{align}
where the exponent results from the WKB approximation and $J_0$ is the only free parameter. From Fig.~\ref{fig:alsi_jc}, we see that this fit matches our $J_c$ values remarkably well;
from the fit, we obtain $J_0 \approx 1 \times 10^{10}$~A/m$^2$.

We can also estimate $J_0$ independently, by starting from Eqs.~\eqref{eqn:jc}--\eqref{eqn:Jtot} and making the following approximations: 
(1) $J_h + J_e \approx J_h$, 
(2) $J_h \approx$~twice $J_h$ without image charge correction, 
(3) rectangular barrier of width $s$ and height $\phi_p$,
(4) $m^*_\mathrm{M} \approx m_0$, 
(5) $\tanh (\Delta(T)/(2k_BT)) \approx 1$ for $T$ sufficiently below the transition temperature, 
and (6) $\phi_p \sim E_g/2$.
With these assumptions and approximations, we obtain:
\begin{align}
    J_0 \approx \frac{\pi e m_0}{h^2} \sqrt{ \frac{E_g}{m^*_h} } ~\frac{\Delta}{s},
    \label{eqn:J0}
\end{align}
For the example shown in Fig.~\ref{fig:alsi_jc}, Eq.~(\ref{eqn:J0}) yields $0.9 \times 10^{10}$~A/m$^2$, essentially the same value as the value obtained from the fit.
Using the approximate $J_0$,
and $\phi_p$ from Table~\ref{tab:alsi_energy_SBH}, 
Eq.~\eqref{eqn:Jc-fit-psbh} reproduces the $J_c$ values from Table~\ref{tab:alsi_jc} within a factor of 2, with a decay constant of $(2/\hbar)\sqrt{2 m^*_h \phi_p} \equiv 2 \kappa_h = 2.2~\mathrm{nm}^{-1}$. 

The rectangular barrier approximation in this case is reasonable because (1) the image-charge corrections are small (without such corrections, the $J_c$'s are only $\sim$2x smaller, as mentioned above); 
(2) the applied voltage $\mathcal{V}$ we use is 
small; and 
(3) the temperature $T$ is low, 20~mK. 
Points \#(1) and \#(2)  make the hole tunneling barrier shaped more like a rectangle than the curved trapezoid illustrated in Fig.~\ref{fig:tunneling}, while 
point \#(3) implies that holes near the Fermi level contribute most, and therefore the effective rectangular barrier height would be $\phi_p$. 

The remarkably good fit obtained here reflects the near-ideal nature of the Al/Si interface.  This contrasts with, for instance, the Al/Al$_2$O$_3$ interface studied in Ref.~\cite{ref:kim}, where equivalent barrier heights (0.043--0.11~eV) much smaller than the physical SBH ($\sim$1--3~eV~\cite{ref:Al-Al2O3-SBH_calc_1, ref:Al-Al2O3-SBH_calc_2}) had to be assumed to match actual calculated transmission coefficients in a fit to a WKB approximation, indicative of the more complicated interfacial structure and greater sensitivity to thickness fluctuations for that junction.

Finally, one might think that for a Si thickness of $\sim$6~nm, quantum confinement effects may play a role in shifting the band extrema, thereby affecting the estimated tunneling currents. First, we note that for this type of metal/semiconductor/metal junction the carrier wavefunctions can freely extend into the metal regions, and thus quantum confinement in principle is not present. Second, even if confinement would occur, the impact on the calculated results would be minor. We performed estimates for the limiting case of an infinite square well of width $\sim$6~nm, assessing the shifts in ground-state energies for light holes, heavy holes, and electrons. 
We found that $J_h$ remains significantly larger than $J_e$, and $J_c$ becomes smaller by less than a factor of three. We conclude that even if quantum confinement were present it would have only a small quantitative impact on our results.


\section{Results: \texorpdfstring{C\MakeLowercase{o}S\MakeLowercase{i}$_2$(111)/S\MakeLowercase{i}(111)}{CoSi2(111)/Si(111)}} \label{sec:results-cosi2}

\subsection{Structures and formation energies} \label{subsec:results-cosi2_struct-en}

\begin{figure*}[t!]
    \centering
    \includegraphics[width=\linewidth]{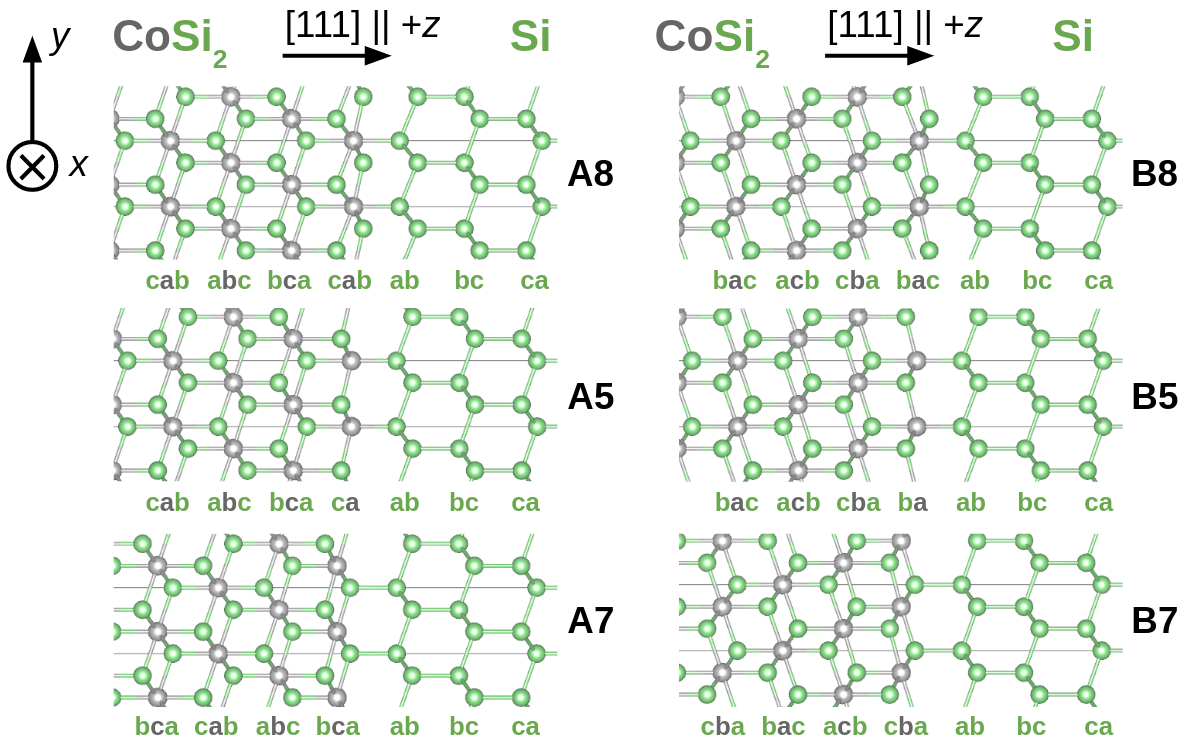}

    \caption{
        The six crystal structures of CoSi$_2$(111)/Si(111) interfaces proposed in the literature, all of which are considered in the present work: A8, B8, A5, B5, A7, and B7. The letter A or B denotes whether CoSi$_2$ and Si have the same (A) or opposite (B) fcc stacking sequences. 
        The number 5, 7, or 8 denotes the coordination of the Co atoms at the interface. 
    }

    \label{fig:cosi2}
\end{figure*}

Because of the relatively close lattice match between CoSi$_2$ and Si, a commensurate interface can be created by matching unit cells of both materials across the interface in a 1$\times$1 structure.
Figure~\ref{fig:cosi2} shows the six structures of CoSi$_2$(111)/Si(111) interfaces that have been proposed in the literature based on experimental observations and theoretical considerations~\cite{ref:gibson, ref:tung-1984, ref:hamann, ref:catana, ref:bulle, ref:stadler, ref:li, ref:seubert}. 
Similar to the case of Al(111)/Si(111), the interface can have type-A and type-B orientations, based on whether the fcc stacking sequence in the metal is the same as or opposite to the stacking sequence in Si. 
For each orientation type, there are three structures, classified by the number of Si atoms bonded to each Co atom at the interface (i.e., the coordination number of the interfacial Co): 5, 7, and 8. 
For A8/B8, the Co coordination number is the same as in the bulk. 
This results in one dangling bond on each Si atom at the interface on the CoSi$_2$ side. 
Structures A5/B5 are created by removing this Si atom from A8/B8. 
Structures A7/B7, finally, are created by starting from A8/B8 and shifting the CoSi$_2$ layer laterally relative to the Si layer, in the process breaking the Co-Si bond across the interface and bonding the resulting Si atom on the Si side with the Si atom that had the dangling bond in A8/B8. 

Matching the in-plane lattice parameter of CoSi$_2$ to that of Si results in an in-plane strain of $\sim$2\% for CoSi$_2$. 
We find that a Brillouin-zone sampling mesh of $12 \times 12 \times 1$ is sufficient to converge the average potential difference $\Delta \bar{V}$ to within 0.01~eV.

We find that allowing all atoms on the CoSi$_2$ to relax along the (111) plane (as opposed to keeping the central layers fixed) does not change the structure, and the resulting energy and $\Delta \bar{V}$ differ by within 3~meV per ($1\times 1$)Si and 5~meV, respectively. This preferential lateral alignment of the Si and CoSi$_2$ is to be expected, since there is a clear covalent bonding pattern across the CoSi$_2$(111)/Si(111) interface as seen in Fig.~\ref{fig:cosi2}.

For all six structures, we relax two CoSi$_2$ trilayers and two Si-Si bilayers near each interface. 
Relaxing three CoSi$_2$ trilayers and three Si-Si bilayers changes $\Delta \bar{V}$ by $\leq$10~meV, as seen in Table~\ref{tab:deltaV_cosi2} in the Supplemental Material~\cite{ref:supplemental}. 
The magnitudes of the relaxations of these layers are reported in Tables \ref{tab:cosi2si_rlx_AB5} and \ref{tab:cosi2si_rlx_AB78}. 
The largest displacements correspond to the atom at the interface on the CoSi$_2$ side. The Si atoms on the Si side of the interface tend to move less; for A7 and B7, they move by only 0.002~{\AA} and 0.004~{\AA}, probably because they bond to Si atoms instead of Co atoms across the interface. 

\begin{table}[h!]
\caption{
    Magnitudes of relaxations of CoSi$_2$(111)/Si(111) layers near the interface for A5 and B5 structures. Because each layer only contains one atom, there is only one displacement value for each layer. Si$_\mathrm{M}$ denotes Si corresponding to CoSi$_2$ layers.
    \label{tab:cosi2si_rlx_AB5}
}
\begin{ruledtabular}
\begin{tabular}{lcc}
    Layer & \multicolumn{2}{c}{Displacement (\AA)} \\
                        & A5    & B5    \\
    \hline
    Co 7 (fixed)        &     0 &     0 \\
    Si$_\mathrm{M}$ 6   & 0.056 & 0.059 \\
    Si$_\mathrm{M}$ 5   & 0.016 & 0.014 \\
    Co 4                & 0.044 & 0.047 \\
    Si$_\mathrm{M}$ 3   & 0.197 & 0.189 \\
    Si$_\mathrm{M}$ 2   & 0.191 & 0.164 \\
    Co 1 (at interface) & 0.260 & 0.243 \\
    Si 1 (at interface) & 0.063 & 0.077 \\
    Si 2                & 0.013 & 0.011 \\
    Si 3                & 0.011 & 0.011 \\
    Si 4                & 0.000 & 0.001 \\
    Si 5 (fixed)        &     0 &     0 \\
\end{tabular}
\end{ruledtabular}
\end{table}

\begin{table}[h!]
\caption{
    Magnitude of relaxations of CoSi$_2$(111)/Si(111) layers near the interface for A7, B7, A8, and B8 structures. 
    \label{tab:cosi2si_rlx_AB78}
}
\begin{ruledtabular}
\begin{tabular}{lcccc}
    Which layer & \multicolumn{4}{c}{Displacement (\AA)} \\
                                     & A7    & B7    & A8    & B8    \\
    \hline
    Si$_\mathrm{M}$ 7 (fixed)        &     0 &     0 &     0 &     0 \\
    Si$_\mathrm{M}$ 6                & 0.000 & 0.000 & 0.009 & 0.006 \\
    Co 5                             & 0.011 & 0.010 & 0.010 & 0.010 \\
    Si$_\mathrm{M}$ 4                & 0.105 & 0.121 & 0.035 & 0.036 \\
    Si$_\mathrm{M}$ 3                & 0.019 & 0.033 & 0.032 & 0.032 \\
    Co 2                             & 0.146 & 0.174 & 0.096 & 0.097 \\
    Si$_\mathrm{M}$ 1 (at interface) & 0.299 & 0.280 & 0.240 & 0.244 \\
    Si 1 (at interface)              & 0.002 & 0.004 & 0.126 & 0.095 \\
    Si 2                             & 0.003 & 0.005 & 0.007 & 0.010 \\
    Si 3                             & 0.001 & 0.001 & 0.007 & 0.002 \\
    Si 4                             & 0.001 & 0.001 & 0.000 & 0.001 \\
    Si 5 (fixed)                     &     0 &     0 &     0 &     0 \\
\end{tabular}
\end{ruledtabular}
\end{table}

Tables \ref{tab:cosi2si_layerSpacing_AB5} and \ref{tab:cosi2si_layerSpacing_AB78} show the interlayer spacings between two adjacent layers. 
We see that the spacings near the fixed layers are quite close to the bulk spacings.
We also observe that the Co-Si distance across the interface is 2.275~{\AA} for A5, 2.288~{\AA} for B5, 2.331~{\AA} for A8, and 2.355~{\AA} for B8,
all of which are close to the Co-Si bond length along the [111] direction of strained bulk CoSi$_2$, 2.308~{\AA}. 
For A7 and B7, the Si-Si distances across the interface are 2.376~{\AA} and 2.382~{\AA}, again very close to the bulk Si-Si bond length, 2.368~{\AA}, and close to the values reported in Ref.~\cite{ref:stadler}, 2.41~{\AA} and 2.40~{\AA}. 
Finally, for A8 and B8, the Si-Si interlayer distances across the interface are 1.898~{\AA} and 1.927~{\AA}, close to the values reported in Ref.~\cite{ref:stadler}, 1.87~{\AA} and 1.91~{\AA}.

\begin{table}[h!]
\caption{
    Interlayer spacings of CoSi$_2$(111)/Si(111) near the interface for A5 and B5 structures.
    \label{tab:cosi2si_layerSpacing_AB5}
}
\begin{ruledtabular}
\begin{tabular}{lcc}
    Which layers & \multicolumn{2}{c}{Interlayer spacing (\AA)} \\
                                                         & A5    & B5    \\
    \hline
    Co 10 $-$ Si$_\mathrm{M}$ 9 (both fixed)             & 0.769 & 0.769 \\
    Si$_\mathrm{M}$ 9 $-$ Si$_\mathrm{M}$ 8 (both fixed) & 1.539 & 1.539 \\
    Si$_\mathrm{M}$ 8 $-$ Co 7 (both fixed)              & 0.769 & 0.769 \\
    Co 7 (fixed) $-$ Si$_\mathrm{M}$ 6                   & 0.714 & 0.710 \\
    Si$_\mathrm{M}$ 6 $-$ Si$_\mathrm{M}$ 5              & 1.578 & 1.584 \\
    Si$_\mathrm{M}$ 5 $-$ Co 4                           & 0.741 & 0.736 \\
    Co 4 $-$ Si$_\mathrm{M}$ 3                           & 0.616 & 0.627 \\
    Si$_\mathrm{M}$ 3 $-$ Si$_\mathrm{M}$ 2              & 1.927 & 1.892 \\
    Si$_\mathrm{M}$ 2 $-$ Co 1                           & 0.319 & 0.362 \\
    Co 1 $-$ Si 1 (across interface)                     & 2.275 & 2.288 \\
    Si 1 $-$ Si 2                                        & 0.839 & 0.856 \\
    Si 2 $-$ Si 3                                        & 2.370 & 2.368 \\
    Si 3 $-$ Si 4                                        & 0.800 & 0.800 \\
    Si 4 $-$ Si 5 (fixed)                                & 2.368 & 2.369 \\
    Si 5 $-$ Si 6 (both fixed)                           & 0.789 & 0.789 \\
    Si 6 $-$ Si 7 (both fixed)                           & 2.368 & 2.368 \\
\end{tabular}
\end{ruledtabular}
\end{table}

\begin{table}[h!]
\caption{
    Interlayer spacings of CoSi$_2$(111)/Si(111) near the interface for A7, B7, A8, and B8 structures.
    \label{tab:cosi2si_layerSpacing_AB78}
}
\begin{ruledtabular}
\begin{tabular}{lcccc}
    Which layers & \multicolumn{4}{c}{Interlayer spacing (\AA)} \\
                                                          & A7    & B7    & A8    & B8    \\
    \hline
    Si$_\mathrm{M}$ 10 $-$ Si$_\mathrm{M}$ 9 (both fixed) & 1.539 & 1.539 & 1.539 & 1.539 \\
    Si$_\mathrm{M}$ 9 $-$ Co 8 (both fixed)               & 0.769 & 0.769 & 0.769 & 0.769 \\
    Co 8 $-$ Si$_\mathrm{M}$ 7 (both fixed)               & 0.769 & 0.769 & 0.769 & 0.769 \\
    Si$_\mathrm{M}$ 7 (fixed) $-$ Si$_\mathrm{M}$ 6       & 1.539 & 1.539 & 1.548 & 1.545 \\
    Si$_\mathrm{M}$ 6 $-$ Co 5                            & 0.780 & 0.780 & 0.770 & 0.773 \\
    Co 5 $-$ Si$_\mathrm{M}$ 4                            & 0.653 & 0.638 & 0.724 & 0.723 \\
    Si$_\mathrm{M}$ 4 $-$ Si$_\mathrm{M}$ 3               & 1.663 & 1.692 & 1.541 & 1.543 \\
    Si$_\mathrm{M}$ 3 $-$ Co 2                            & 0.605 & 0.563 & 0.897 & 0.898 \\
    Co 2 $-$ Si$_\mathrm{M}$ 1                            & 0.616 & 0.663 & 0.433 & 0.428 \\
    Si$_\mathrm{M}$ 1 $-$ Si 1 (across interface)         & 2.376 & 2.382 & 1.898 & 1.927 \\
    Si 1 $-$ Si 2                                         & 0.784 & 0.788 & 0.922 & 0.894 \\
    Si 2 $-$ Si 3                                         & 2.369 & 2.372 & 2.368 & 2.360 \\
    Si 3 $-$ Si 4                                         & 0.789 & 0.790 & 0.783 & 0.788 \\
    Si 4 $-$ Si 5 (fixed)                                 & 2.369 & 2.369 & 2.368 & 2.367 \\
    Si 5 $-$ Si 6 (both fixed)                            & 0.789 & 0.789 & 0.789 & 0.789 \\
    Si 6 $-$ Si 7 (both fixed)                            & 2.368 & 2.368 & 2.368 & 2.368 \\
\end{tabular}
\end{ruledtabular}
\end{table}

Table \ref{tab:cosi2si_energy} shows the interface formation energies per $(1 \times 1)$ Si for all considered CoSi$_2$(111)/Si(111) structures. 
Note that, unlike Al(111)/Si(111), the energies vary significantly for different structures. 
We attribute this to the more covalent character of the bonding across the interface, leading to 
different atomic structures (Fig.~\ref{fig:cosi2}) and layer-projected densities-of-states (Fig.~\ref{fig:lpdos_cosi2si} in the Supplemental Material~\cite{ref:supplemental}). 
Similar to Refs.~\onlinecite{ref:hamann} and \onlinecite{ref:stadler}, we find the eightfold-coordinated structures to be lowest in energy and the fivefold-coordinated structures to be least energetically favorable, and therefore the latter may be very difficult to grow. 
We note that the energies in \cite{ref:hamann} are larger than both ours and those reported in \cite{ref:stadler}, likely because no atomic relaxation was allowed. 

\begin{table}[h!]
\caption{
    Interface formation energies per (1 $\times$ 1) in-plane Si ($\gamma^\text{f} A_{(1 \times 1) \mathrm{Si}}$) for all CoSi$_2$(111)/Si(111) structures considered. 
    \label{tab:cosi2si_energy}
}
\begin{ruledtabular}
\begin{tabular}{lcccccc}
    & \multicolumn{6}{c}{$\gamma^\text{f} A_{(1 \times 1) \mathrm{Si}}$ (eV)} \\
                                        & A5   & B5   & A7   & B7   & A8   & B8 \\
    \hline
    This work, PBE                      & 1.90 & 1.94 & 0.60 & 0.66 & 0.40 & 0.37 \\
    This work, HSE                      & -    & -    & -    & -    & 0.42 & 0.41 \\
    Ref.~\onlinecite{ref:hamann} (LDA)  & 2.20 & -    & 0.86 & -    & 0.68 & 0.53 \\
    Ref.~\onlinecite{ref:stadler} (GGA) & -    & -    & 0.45 & 0.49 & 0.39 & 0.43 \\
    \end{tabular}
\end{ruledtabular}
\end{table}

For the lowest-energy structures, A8 and B8, we also performed full HSE calculations, starting from the relaxed PBE geometry but scaled to match the HSE lattice parameters and bond lengths. 
We then allowed for atomic relaxation of the atoms near the interface. 
Due to the high computational expense, we stopped the HSE calculations when $\Delta \bar{V}$ values were converged to within 0.01~eV. 
We find that the magnitudes of relaxations and the interlayer spacings differ by as large as 0.04~{\AA} from the PBE ones when normalized to HSE bulk bond lengths. 
As shown in Table \ref{tab:cosi2si_energy}, for A8 and B8 structures, the fully HSE formation energies are merely 0.02~eV and 0.04~eV higher than our PBE formation energies.

\subsection{Schottky barrier heights} 
\label{subsec:results-cosi2_struct-sbh}

Our calculated $p$-type SBHs for all six CoSi$_2$(111)/Si(111) structures are presented in Table~\ref{tab:cosi2si_SBH}. 
The SBHs display variations by as much as $\sim$0.4~eV, which is not surprising, given the distinct differences in bonding. 
Note that the SBHs for A5 and B5 are within 0.01~eV from each other, yet the $p$-SBH for A7 is 0.14~eV higher than that of B7, and the $p$-SBH for A8 is 0.23~eV higher than that of B8. 
We suggest this is because, as seen in Fig.~\ref{fig:cosi2}, 
(1) the lateral positions of the Si atoms with a dangling bond relative to the atoms in the Si slab are different for A8 and B8, and (2) the lateral positions of interfacial Co atoms relative to the atoms in the Si slab are different for A7 and B7, yet (3) for A5 and B5, there are no Si atoms with a dangling bond, and the interfacial Co atoms remain in the same lateral positions relative to the Si slab. 
We also note that for NiSi$_2$(111)/Si(111), a similar interface, experimental measurements have shown that that the $p$-SBH of the A7 structure is 0.14~eV higher than that of B7~\cite{ref:tung_nisi2}. 

As to why A8 has the highest $p$-SBH, we speculate this might be due to the interaction between the Si atoms with dangling bond and the second Si layer away from the interface on the Si side, both of which are at the same in-plane positions. 
This interaction could increase the charge at the dangling-bond site, thereby increasing the electrostatic potential energy on the metal side and ultimately the $p$-SBH. 
The positioning mentioned above is not present in B8, and all other structures do not have Si atoms with a dangling bond.

\begin{table}[h!]
\caption{
    \textit{p}- and \textit{n}-type Schottky barrier heights ($\phi_p$ and $\phi_n$) for all CoSi$_2$(111)/Si(111) structures considered. The values of $\phi_p$ in the literature are also included. 
    Results marked with asterisks (*) use the layer-projected density-of-states method; the others use the potential-alignment method.
    The notation ``PBE+HSE'' indicates that interface calculations are performed with PBE and combined with HSE calculations for bulk as described in Sec.~\ref{sec:approach}.
    \label{tab:cosi2si_SBH}
}
\begin{ruledtabular}
\begin{tabular}{lcccccc}
    & \multicolumn{6}{c}{$\phi_p$ (eV)} \\
                                         & A5   & B5   & A7   & B7   & A8   & B8 \\
    \hline
    This work, PBE+HSE                   & 0.41 & 0.41 & 0.68 & 0.54 & 0.79 & 0.56 \\ 
    This work, HSE                       & -    & -    & -    & -    & 0.87 & 0.42 \\
    This work, PBE                       & 0.06 & 0.06 & 0.33 & 0.18 & 0.43 & 0.20 \\
    Ref.~\onlinecite{ref:stadler} (GGA)  & -    & -    & 0.66 & 0.48 & 0.51 & 0.28 \\
    Ref.~\onlinecite{ref:zhao} (LDA)*    & -    & -    & -    & -    & -    & 0.28 \\
    Ref.~\onlinecite{ref:gao} (LDA+HSE)* & -    & -    & -    & -    & -    & 0.45 \\
    Ref.~\onlinecite{ref:wasey} (PBE)*   & -    & -    & 0.34 & 0.18 & 0.38 & 0.14 \\
    \hline
    \hline
    & \multicolumn{6}{c}{$\phi_n$ (eV)} \\
    \hline
    This work, PBE+HSE                   & 0.76 & 0.76 & 0.48 & 0.63 & 0.38 & 0.61 \\ 
\end{tabular}
\end{ruledtabular}
\end{table}

Table~\ref{tab:cosi2si_SBH} also reports the $p$-type SBHs for the all-HSE calculations of A8 and B8 described in the previous section. 
While the value for B8 is lower by 0.14~eV from our PBE+HSE calculation, $\phi_p$ for A8 is higher by 0.08~eV. 
We speculate that the HSE functional strengthens the effect that makes $\phi_p$ for A8 higher, possibly due to the aforementioned interaction between the Si with a dangling bond and the second Si layer away from the interface on the silicon side. 
In particular, HSE tends to localize electrons more strongly than PBE, and therefore the charge at the dangling-bond site could be more localized on the metal side, leading to higher $p$-SBH. 

The difference in the PBE+HSE and all-HSE results may also contain a contribution from the difference between the in-plane strain of CoSi$_2$ for HSE and that for PBE. 
Within HSE, we find that changing the in-plane strain of CoSi$_2$ from the PBE value (+2.1\%) to the HSE value (+2.6\%) decreases the bulk Fermi level by $\sim$0.1~eV. This may contribute to the 0.14-eV discrepancy of $\phi_p$ for B8 between PBE+HSE and all-HSE.

We now compare our $p$-SBH results with previous first-principles calculations.
As seen in Table~\ref{tab:cosi2si_SBH}, our PBE+HSE results differ from previous calculations by as much as 0.4~eV. 
Nevertheless, the {\it differences} between type-A and type-B orientations agree with both Refs. \onlinecite{ref:stadler} and \onlinecite{ref:wasey}: the $p$-SBH of A7 is $\sim$0.15~eV higher than that of B7, and the $p$-SBH of A8 is $\sim$0.25~eV higher than that of B8, consistent with the expectation that the differences in SBHs are more accurate than their absolute values. 

To investigate the effects of using different functionals, in the table we also include our PBE results of $p$-SBH, calculated entirely using PBE.
We find that, compared to PBE, using HSE functional and structure (with in-plane strain corresponding to the PBE value) increases the Fermi level by 0.18~eV and lowers the Si VBM by 0.17~eV; this explains why our PBE+HSE results are around 0.35~eV higher than our PBE results. 
Our PBE results are within 0.01~eV of the results of \cite{ref:wasey} for A7 and B7, and within 0.06~eV for A8 and B8. 
Our PBE results also agree with Ref.~\onlinecite{ref:stadler} within their quoted uncertainty of 0.2~eV, except for A7 and B7; the deviations might be due to the differences in the GGA functional and pseudopotentials used. 
For B8, our PBE+HSE result is within 0.02~eV of the LDA+HSE result from \cite{ref:gao}. 

Experimentally measured $n$-type SBHs are in the range 0.62--0.9~eV \cite{ref:tung-1984, ref:rosencher, ref:sirringhaus, ref:meyer}. Using the Si band gap of 1.17~eV (at 0 K), our calculated $n$-SBHs are in the range 0.38--0.76~eV as seen in Table~\ref{tab:cosi2si_SBH}. 
Our calculated values mostly overlap with the experimental range.
The experimental papers do not associate the measured SBHs with specific structures, except for Ref.~\onlinecite{ref:tung-1984}, which measured an $n$-SBH of around 0.64~eV at room temperature for a structure that was stated to be consistent with B5. 
This value is lower than our calculated $n$-SBH for B5, which is 0.76~eV. The discrepancy may be due to defects such as misfit dislocations, which Refs.~\cite{ref:tung-1984, ref:rosencher, ref:sirringhaus, ref:meyer} mentioned were present in their samples.

\subsection{Josephson currents} 
\label{subsec:results-cosi2-current}

Using the SBHs from our PBE+HSE calculations, Table~\ref{tab:cosi2si_jc} presents estimates of the Josephson critical current densities, Eq.~\eqref{eqn:jc}, for various Si barrier thicknesses in the Josephson junction. 
Without image-charge corrections, the current densities are $\sim$$2\times$ smaller.
For each thickness, the critical current densities are very similar for A5 and B5 because their SBHs are within 0.01~eV from each other. 
The current densities for A5 and B5 are higher than those for other structures, which we attribute to the $p$-SBHs of A5 and B5 being smaller than those for the other structures, leading to higher hole currents. 
However, as seen in Table~\ref{tab:cosi2si_energy}, these structures have significantly higher interface formation energies than the other structures, and therefore may be very difficult to grow experimentally. 

\begin{table}[h!]
\caption{
    Estimates of Josephson critical current densities ($J_c$) and qubit resonant frequencies ($f_q$) for CoSi$_2$(111)/Si(111) junctions using our PBE+HSE SBHs, for various Si barrier thicknesses $s$. Also listed are the decay constants $2 \kappa_h \equiv (2/\hbar) \sqrt{2 m^*_h \phi_p}$ from fitting $J_c$ to Eq.~\eqref{eqn:Jc-fit-psbh}.
    \label{tab:cosi2si_jc}
}
\begin{ruledtabular}
\begin{tabular}{rrrrrrr}
    $s$ (nm) & \multicolumn{6}{c}{$J_c$ (A/m$^2$)} \\
       & A5     & B5     & A7    & B7     & A8     & B8 \\
    \hline
   5 & $9 \times 10^{  5}$ & $9 \times 10^{  5}$ & $7 \times 10^{  4}$ & $3 \times 10^{  5}$ & $4 \times 10^{  4}$ & $2 \times 10^{  5}$ \\
   6 & $1 \times 10^{  5}$ & $1 \times 10^{  5}$ & $6 \times 10^{  3}$ & $3 \times 10^{  4}$ & $3 \times 10^{  3}$ & $2 \times 10^{  4}$ \\
   7 & $2 \times 10^{  4}$ & $2 \times 10^{  4}$ & $4 \times 10^{  2}$ & $3 \times 10^{  3}$ & $2 \times 10^{  2}$ & $2 \times 10^{  3}$ \\
  10 & 46 & 46 & 0.2 & 3.5 & 0.05 & 2.4 \\
  20 & $10^{ -7}$ & $10^{ -7}$ & $10^{-12}$ & $10^{- 9}$ & $10^{-13}$ & $10^{-11}$ \\
    \hline
    \hline
    & \multicolumn{6}{c}{$f_q$ (GHz)} \\
    \hline
   5 & 59 & 59 & 17 & 31 & 12 & 29 \\
   6 & 23 & 23 & 5.0 & 11 & 3.4 & 9.8 \\
   7 & 9.3 & 9.3 & 1.5 & 3.8 & 0.9 & 3.4 \\
  10 & 0.6 & 0.6 & 0.04 & 0.2 & 0.02 & 0.1 \\
  20 & $10^{ -4}$ & $10^{ -4}$ & $10^{ -7}$ & $10^{ -6}$ & $10^{ -8}$ & $10^{ -6}$ \\
    \hline
    \hline
    & \multicolumn{6}{c}{$2 \kappa_h$ (nm$^{-1}$)} \\
    \hline
    & 1.86 & 1.86 & 2.40 & 2.12 & 2.57 & 2.16 \\
\end{tabular}
\end{ruledtabular}
\end{table}

As discussed in Sec.~\ref{subsec:tunneling}, for FinMET devices the desired qubit frequency $f_q$ is 4--5~GHz. 
From our estimations, all structures potentially satisfy this expectation at semiconductor thicknesses in the range of 5--10~nm.
Among these structures, B8 has the lowest interface formation energy, as seen in Table~\ref{tab:cosi2si_energy}, and therefore may be the easiest to grow.

\begin{figure}[h!]
    \centering
    \includegraphics[width=\linewidth]{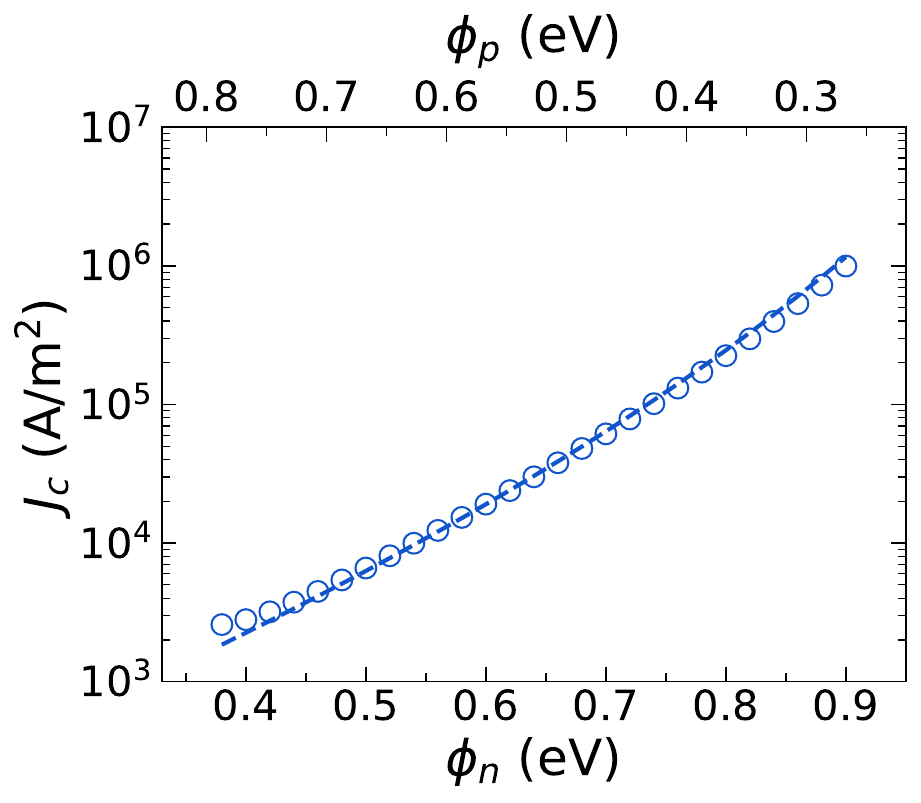}

    \caption{
        Estimated $J_c$ as a function of SBH of CoSi$_2$(111)/Si(111) at Si thickness $s$ = 6~nm (circles) along with the fit to the WKB tunneling probability formula of holes tunneling through a rectangular barrier of height $\phi_p$ (dashed curve).
    }

    \label{fig:cosi2si_jc}
\end{figure}

Finally, we analyze how $J_c$ depends on the SBHs. Figure~\ref{fig:cosi2si_jc} shows our calculated $J_c$ for $\phi_n$ in the range 0.38--0.9~eV, which includes both our calculated and experimentally measured n-SBHs as discussed in the previous section. 
Equation~\eqref{eqn:Jc-fit-psbh} yields a remarkably good fit to our $J_c$ data. 
Similar to the case of Al(111)/Si(111), here we also find that 
(1) hole tunneling dominates ($J_h \sim 30$--$10^4$~A/m$^2$, while $J_e \sim 10^{-3}$--$10$~A/m$^2$), and
(2) image-charge corrections effects are small. 
The remarkably good fit reflects the near-ideal nature of the CoSi$_2$/Si interfaces. 
The coefficient $J_0$ from the fit is $0.9 \times 10^{10}$~A/m$^2$, approximately equal to the expression given by Eq.~\eqref{eqn:J0}. 
Using this approximate $J_0$ and the PBE+HSE $\phi_p$ values from Table~\ref{tab:cosi2si_SBH}, Eq.~\eqref{eqn:Jc-fit-psbh} reproduces the $J_c$ values from Table~\ref{tab:cosi2si_jc} within a factor of 2, with decay constants $2 \kappa_h$ shown in the table.

\FloatBarrier 


\section{Conclusions} \label{sec:conclusion}

This work presents first-principles calculations of structural relaxations, formation energies, and Schottky barrier heights (SBHs) for Al(111)/Si(111) and CoSi$_2$(111)/Si(111) interfaces. 
For Al(111)/Si(111), we find that due to the metallic nature of the bonding, the interface energies and SBHs are not sensitive to the details of the structure, including lateral displacements 
of the Al layers relative to the Si layers within the interfacial structure of ($4 \times 4$)Al matched to ($3 \times 3$)Si, which accommodates the lattice mismatch.
For CoSi$_2$(111)/Si(111), on the other hand, the covalent nature of the bonding leads to distinct energetic and structural differences between the six possible interface structures we considered, also resulting in different electronic properties and SBHs.
Fivefold-coordinated structures are energetically unfavorable, while eightfold-coordinated structures are lowest in energy.
The SBHs for the fivefold-coordinated structures are very similar, while for the sevenfold- and eightfold-coordinated structures, the $p$-type SBHs are higher for the A orientation than for the B orientation. 

We also provide estimates of the Josephson critical currents for the calculated interfaces, for various tunneling barrier thicknesses.
These values are then used to estimate qubit resonance frequencies for FinMETs, demonstrating that qubit frequencies of 4--5~GHz can be obtained with Si barrier thicknesses around 5--10 nm.
A fit of the critical current results to a model based on the WKB tunneling probability for a rectangular barrier shows predictive capability for the change in current (and frequency) as a function of barrier height.
The results should be useful as a guide in developing novel silicon-based merged-element transmons. 


\begin{acknowledgements}

We gratefully acknowledge discussions with 
R. W. Simmonds, A. P. McFadden, D. Waldhoer, and C. A. Broderick. 
This work was supported by the Army Research Office (grant number W911NF-22-1-0052), and
used the SDSC Expanse at the University of California, San Diego through allocation DMR070069 from the Advanced Cyberinfrastructure Coordination Ecosystem: Services \& Support (ACCESS) program~\cite{ref:access}, which is supported by National Science Foundation grants \#2138259, \#2138286, \#2138307, \#2137603, and \#2138296.

\end{acknowledgements}


\bibliography{refs} 

\end{document}


\floatsetup[figure]{style=plain,subcapbesideposition=top} 
\floatsetup[table]{style=plain, capposition=top} 

\preprint{APS/123-QED}

\title{Supplemental Material: First-principles studies of Schottky barriers and tunneling properties at Al(111)/Si(111) and CoSi$_2$(111)/Si(111) interfaces}


\author{J. K. Nangoi}
\email[Corresponding author: nangoi@ucsb.edu]{}
\affiliation{
    Materials Department, University of California, Santa Barbara, California 93106-5050, USA
}

\author{C. J. Palmstr{\o}m}
\affiliation{
    Materials Department, University of California, Santa Barbara, California 93106-5050, USA
}

\author{C. G. Van de Walle}
\affiliation{
    Materials Department, University of California, Santa Barbara, California 93106-5050, USA
}


%



\maketitle



\begin{table}[h!]
\caption{
    Convergence of $\Delta \bar{V}$ for Al(111)/Si(111). 
    ``10 Al,  9 Si-Si, relax 2 sets'' means the supercell contains 10 Al layers and 9 Si-Si bilayers, and two Al layers and two Si-Si bilayers near each interface are relaxed. 
    \label{tab:deltaV_al}
}
\begin{ruledtabular}
\begin{tabular}{ccc}
    Thickness & \multicolumn{2}{c}{$\Delta \bar{V}$ (eV)} \\
                             & A         & B         \\
    \hline
    10 Al,  9 Si-Si, relax 2 sets & $-1.9841$ & $-1.9878$ \\
    13 Al, 12 Si-Si, relax 2 sets & $-1.9460$ & $-1.9494$ \\
    13 Al, 12 Si-Si, relax 3 sets & $-1.9449$ & $-1.9511$ \\
    16 Al, 15 Si-Si, relax 2 sets & $-1.9609$ & $-1.9585$ \\
\end{tabular}
\end{ruledtabular}
\end{table}

\begin{table*}[t!]
\caption{
    Convergence of $\Delta \bar{V}$ for CoSi$_2$(111)/Si(111). 
    ``10 Si-Co-Si,  9 Si-Si, relax 2 sets'' means the supercell contains 10 Si-Co-Si trilayers and 9 Si-Si bilayers, and two CoSi$_2$ trilayers and two Si-Si bilayers near each interface are relaxed. 
    \label{tab:deltaV_cosi2}
}
\begin{ruledtabular}
\begin{tabular}{ccccccc}
    Thickness & \multicolumn{6}{c}{$\Delta \bar{V}$ (eV)} \\
                             & A5        & B5        & A7        & B7        & A8        & B8 \\
    \hline
    10 Si-Co-Si,  9 Si-Si, relax 2 sets & -         & -         & -         & -         & $-3.5701$ & $-3.7912$ \\
    10 Si-Co-Si,  9 Si-Si, relax 3 sets & -         & -         & -         & -         & $-3.5765$ & - \\
    11 Si-Co-Si,  9 Si-Si, relax 2 sets & -         & -         & $-3.6754$ & -         & -         & - \\
    11 Si-Co-Si,  9 Si-Si, relax 3 sets & -         & -         & $-3.6780$ & -         & -         & - \\
    13 Si-Co-Si, 12 Si-Si, relax 2 sets & -         & -         & -         & -         & $-3.5756$ & $-3.8038$ \\
    13 Si-Co-Si, 12 Si-Si, relax 3 sets & -         & -         & -         & -         & -         & $-3.7940$ \\
    14 Si-Co-Si, 12 Si-Si, relax 2 sets & -         & -         & $-3.6807$ & $-3.8228$ & -         & - \\
    14 Si-Co-Si, 12 Si-Si, relax 3 sets & -         & -         & -         & $-3.8161$ & -         & - \\
    16 Si-Co-Si, 12 Si-Si, relax 2 sets & -         & -         & -         & -         & -         & $-3.7986$ \\
    Co-Si + 14 Si-Co-Si + Si-Co, 15 Si-Si, relax 2 sets & $-3.9646$ & $-3.9626$ & -         & -         & -         & - \\
    17 Si-Co-Si, 15 Si-Si, relax 2 sets & -         & -         & -         & $-3.8375$ & -         & - \\
    Co-Si + 17 Si-Co-Si + Si-Co, 18 Si-Si, relax 2 sets & $-3.9479$ & $-3.9483$ & -         & -         & -         & - \\
    Co-Si + 17 Si-Co-Si + Si-Co, 18 Si-Si, relax 3 sets & $-3.9486$ & $-3.9473$ & -         & -         & -         & - \\
    Co-Si + 20 Si-Co-Si + Si-Co, 21 Si-Si, relax 2 sets & $-3.9496$ & $-3.9390$ & -         & -         & -         & - \\
\end{tabular}
\end{ruledtabular}
\end{table*}

\begin{table}[h!]
\caption{
    Chosen layer thicknesses and differences of $\Delta \bar{V}$ with respect to thicker layers, $|\delta \Delta \bar{V}|$, for Al(111)/Si(111) and CoSi$_2$(111)/Si(111). All are ``relax 2 sets.'' 
    \label{tab:chosenLayerThicknesses}
}
\begin{ruledtabular}
\begin{tabular}{lll}
    Structure & Chosen thickness & $|\delta \Delta \bar{V}|$ (eV) \\
    \hline
    A  & 13 Al, 12 Si-Si       & $<0.015$ \\
    B  & 13 Al, 12 Si-Si       & $<0.01$  \\
    A5 & Co-Si + 17 Si-Co-Si + Si-Co, 18 Si-Si & $<0.01$  \\
    B5 & Co-Si + 17 Si-Co-Si + Si-Co, 18 Si-Si & $<0.01$  \\
    A7 & 11 Si-Co-Si,  9 Si-Si & $<0.01$  \\
    B7 & 14 Si-Co-Si, 12 Si-Si & $<0.015$ \\
    A8 & 10 Si-Co-Si,  9 Si-Si & $<0.01$  \\
    B8 & 13 Si-Co-Si, 12 Si-Si & $<0.01$  \\
\end{tabular}
\end{ruledtabular}
\end{table}

\begin{figure*}[b!]
    \centering
    \sidesubfloat[]{\includegraphics[width=0.49\linewidth]{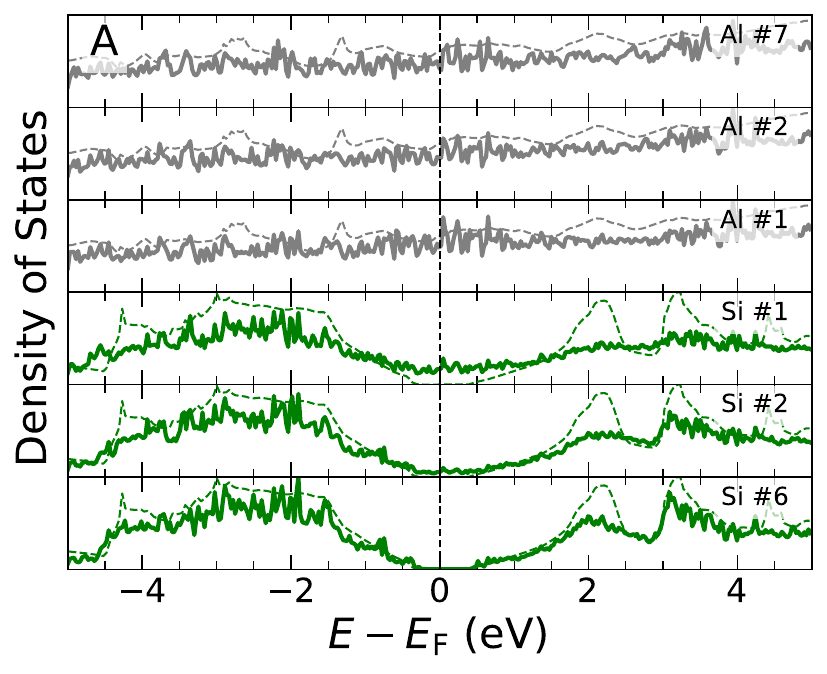}\label{fig:LPDOS_AlSi_A}}
    \sidesubfloat[]{\includegraphics[width=0.46\linewidth]{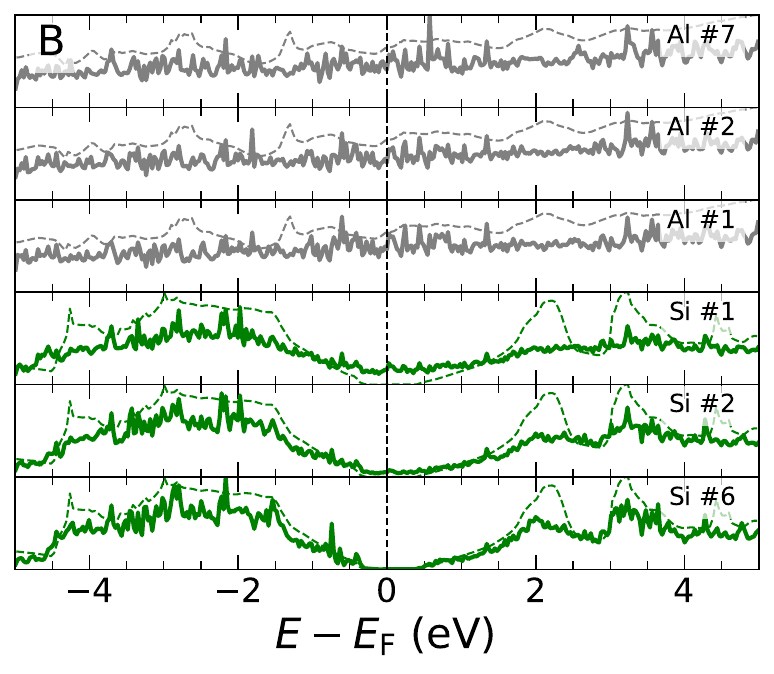}\label{fig:LPDOS_AlSi_B}}%

    \caption{
        Density-of-states (DOS) from PBE calculations of Al(111)/Si(111) (a) type A and (b) type B. For each plot, Brillouin zone sampling mesh of $10 \times 10 \times 1$ is used. Gray curves correspond to Al, green to Si. Solid curves are layer-projected DOS, dashed curves are bulk DOS. 
        ``Al \#1'' means the first Al layer from the interface, and ``Si \#1'' means the first Si-Si bilayer from the interface. The highest numbers correspond to the central layers.
        For each element, the layer-projected DOS curves are scaled so that the maximum over all of these curves is on the top of the vertical axis. For bulk Al DOS, the maximum of the curve is set to equal the maximum of layer-projected DOS in the central Al layer. 
        Likewise for bulk Si DOS.
    }

    \label{fig:lpdos_alsi}
\end{figure*}

\begin{figure*}[b!]
    \centering
    \sidesubfloat[]{\includegraphics[width=0.49\linewidth]{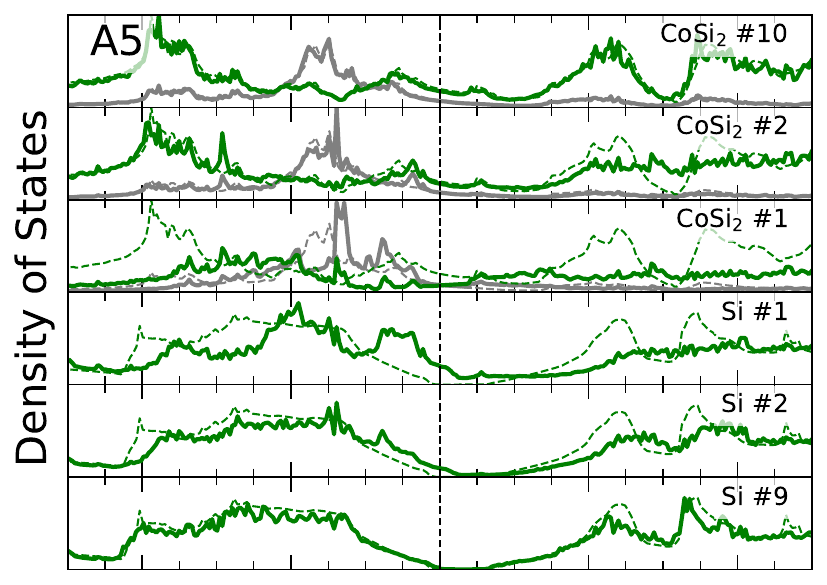}\label{fig:LPDOS_CoSi2Si_A5}}
    \sidesubfloat[]{\includegraphics[width=0.46\linewidth]{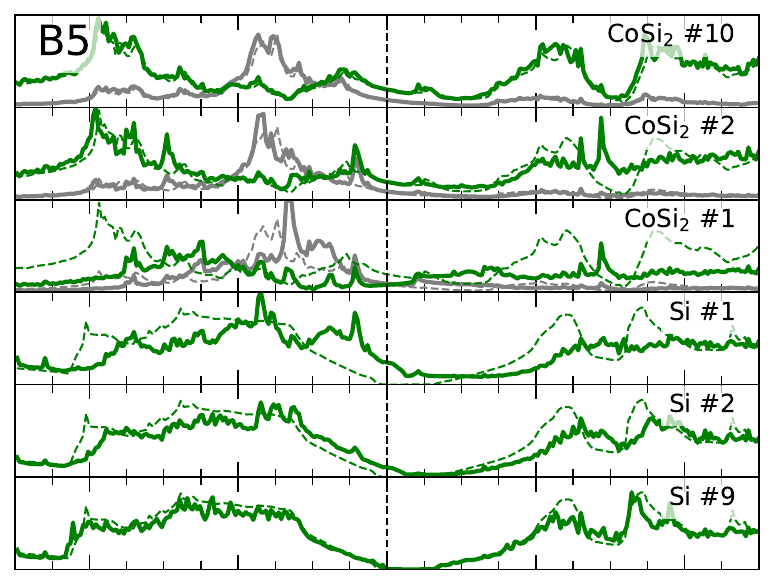}\label{fig:LPDOS_CoSi2Si_B5}}%
    \\
    \sidesubfloat[]{\includegraphics[width=0.49\linewidth]{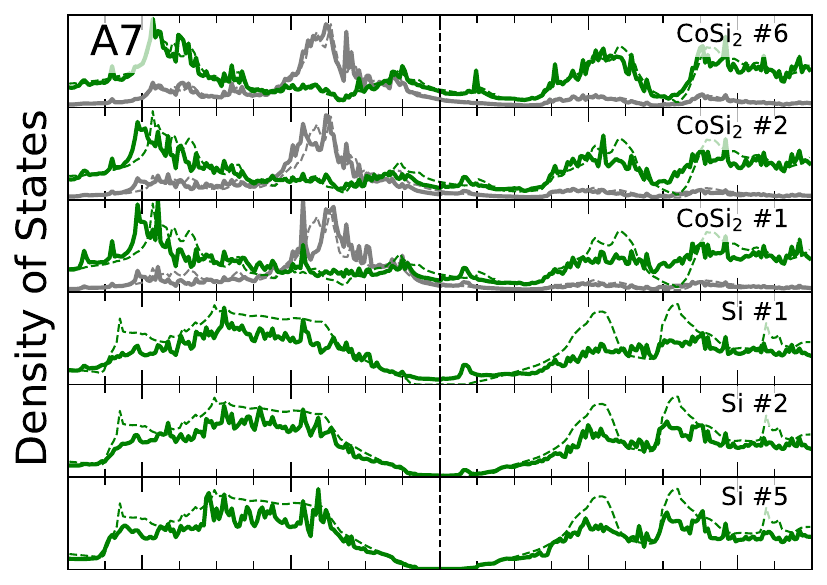}\label{fig:LPDOS_CoSi2Si_A7}}
    \sidesubfloat[]{\includegraphics[width=0.46\linewidth]{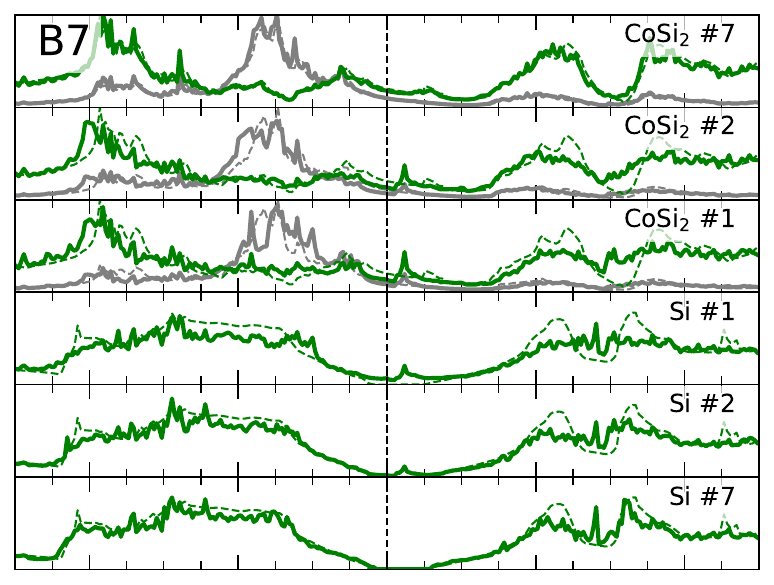}\label{fig:LPDOS_CoSi2Si_B7}}%
    \\
    \sidesubfloat[]{\includegraphics[width=0.49\linewidth]{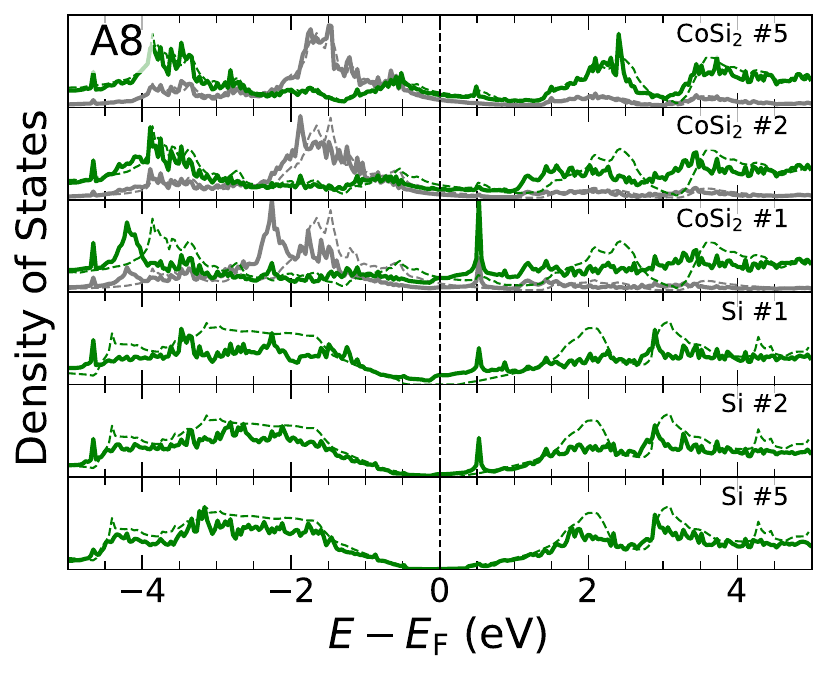}\label{fig:LPDOS_CoSi2Si_A8}}
    \sidesubfloat[]{\includegraphics[width=0.46\linewidth]{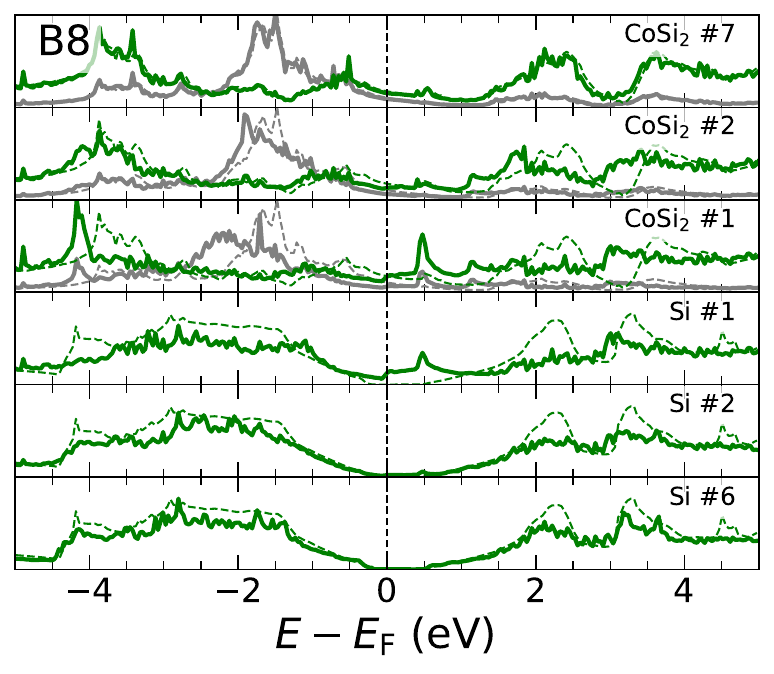}\label{fig:LPDOS_CoSi2Si_B8}}%
    \\

    \caption{
        Density-of-states (DOS) from PBE calculations of CoSi$_2$(111)/Si(111) structure (a) A5, (b) B5, (c) A7, (d) B7, (e) A8, and (f) B8. For each plot, Brillouin zone sampling mesh of $24 \times 24 \times 1$ is used. Gray curves correspond to Co, green to Si. Solid curves are layer-projected DOS, dashed curves are bulk DOS. 
        ``CoSi$_2$ \#1'' means the first CoSi$_2$ trilayer from the interface.
        For each element, the layer-projected DOS curves are scaled so that the maximum over all of these curves is on the top of the vertical axis. For bulk CoSi$_2$ DOS, the maximum of the curve of each element is set to equal the maximum of layer-projected DOS of the corresponding element in the central CoSi$_2$ trilayer.
    }

    \label{fig:lpdos_cosi2si}
\end{figure*}

\FloatBarrier 
